High-energy cosmic rays and the Greisen-Zatsepin-Kuz'min effect

A A Watson
School of Physics and Astronomy
University of Leeds
Leeds LS2 9JT

E-mail: a.a.watson@leeds.ac.uk

"All things keep on, in everlasting motion. Out of the infinite come the particles speeding above, below, in endless dance"
Lucretius, 'De Rerum Natura' 50 BC

**Abstract:** Although cosmic rays were discovered over 100 years ago their origin remains uncertain. They have an energy spectrum that extends from ~ 1 GeV to beyond $10^{20}$ eV, where the rate is less than 1 particle per km$^2$ per century. Shortly after the discovery of the cosmic microwave background in 1965, it was pointed out that the spectrum of cosmic rays should steepen fairly abruptly above about 4 x $10^{19}$ eV, provided the sources are distributed uniformly throughout the Universe. This prediction, by Greisen and by Zatsepin and Kuz'min, has become known as the GZK-effect and in this article I discuss the current position with regard to experimental data on the energy spectrum of the highest cosmic-ray energies that have been accumulated in a search that has lasted nearly 50 years. Although there is now little doubt that a suppression of the spectrum exists near the energy predicted, it is by no means certain that this is a manifestation of the GZK-effect as it might be that this energy is also close to the maximum to which sources can accelerate particles, with the highest-energy beam containing a large fraction of nuclei heavier than protons. The way forward is briefly mentioned.

**1. Introduction**
1.1 Some background information about cosmic rays
A detailed understanding of the properties and origin of cosmic rays with energies greater than 1 Joule (6.3 x $10^{18}$ eV) remains lacking over 50 years after their discovery [1]. Following the first claim for an event above ~ $10^{20}$ eV made by Linsley in 1963 using observations at Volcano Ranch [2], painstaking work by dedicated groups resulted in reports of detections of a small number of events above this energy. In this article the methods devised for detecting cosmic rays above $10^{19}$ eV where the integral flux is below 1 per km$^2$ per steradian per year (km$^{-2}$ sr$^{-1}$ yr$^{-1}$ in what follows) will be described and the measurements made of the energy spectrum will be discussed in some detail. It will be shown that there is convincing evidence that the cosmic-ray flux is suppressed near to 4 x $10^{19}$ eV although it remains unclear as to whether this is the long-sought Greisen-Zatsepin-Kuz'min effect, and thus an effect of cosmic-ray propagation through the Universe, or whether it is an indication of the limit to which high-energy particles can be accelerated: either conclusion is of astrophysical importance. While these particles are particularly fascinating as they represent matter in its most extreme departure from thermal equilibrium, it is appropriate in an article of this type to start with some background about cosmic rays of lower energy and the cosmic-ray phenomenon in general.

The existence of a high-energy radiation stream incident continuously on the earth was discovered because of the adventurous spirit of the Austrian physicist, Victor Hess. Through a series of daring balloon flights, culminating in his famous flight of 7 August 1912, Hess established that the ionisation of air at 5000 m was more than twice that at sea-level. The name 'cosmic rays', for what we now recognise as mainly charged particles coming from sources somewhere beyond the solar system, is often credited as having been coined by Robert Millikan in the 1920s although German-speaking physicists commonly used the phrase *'kosmische strahlung'* much earlier.  It took nearly 30 years from 1912 to understand that the incoming radiations were not the gamma-rays that Millikan in particular had supposed but were largely protons and, at least at low energies, cosmic rays are now known to encompass the nuclei of all of the stable and extremely long-lived elements.  Schein et al., [3] finally established that the proton dominance, putting to rest the idea the mainly positively-charged nature of the primaries, determined from geomagnetic studies, was indicative of positrons as the main species.

Until the early 1950s the field of cosmic rays was a rich hunting ground for what came to be called 'elementary particles'.  The positron, muon, the charged pions and several of the 'strange particles' were all discovered in cosmic rays; two of the giants of UK physics of the last century, C F Powell and P M S Blackett, won their Nobel Prizes for work in this field.  From an astrophysical viewpoint an important outcome of cosmic-ray studies was the postulate of Fermi that charged particles could be accelerated by clouds of magnetised gas moving within our galaxy.  He created this model in part to counter the contention of Teller and Richtmeyer and of Alfvén that cosmic rays come dominantly from the Sun.  While we now know that Fermi's original mechanism is too slow to be effective, it was adapted in the late 1970s in 'first-order' form, usually known as diffusive shock acceleration, as a way to accelerate charge particles efficiently at shock fronts.  Evidence for the first-order mechanism is now available directly from instruments on spacecraft deployed in the inter-planetary medium and it has been inferred that the process will operate in many shock environments.

By 1953 it was apparent that particle physics was best pursued at accelerators and many who had been 'cosmic-ray physicists' left the field to work at national and international high-energy physics laboratories.  Those who remained used cosmic rays to study nuclear interactions at energies unobtainable in man-made accelerators and to search for primary particles of greater and greater energy with the expectation that above some high energy the particles would travel nearly rectilinearly through the intervening magnetic fields and so reveal their sources.

In many books on high-energy astrophysics a plot of the cosmic-ray energy spectrum similar to that of figure 1 [4] will be found: the flux falls by 25 orders of magnitude over 11 decades of energy.  By comparison with a plot of the electromagnetic spectrum over the same span, this spectrum is remarkably featureless.  There is an increase of slope at about $3 - 5 \times 10^{15}$ eV, known as the 'knee', where the flux of particles is about one per square metre per year.  The spectrum flattens again near $\sim 4 \times 10^{18}$ eV, the 'ankle', above which only a few particles arrive per square kilometre per year.  Recently a second knee in the spectrum of particles with greater than average mass has been reported at $\sim 8 \times 10^{16}$ eV, about 26 times higher than the knee and thus consistent with the idea that knee features are rigidity-related [5].  In this article I will concentrate on the higher-energy region where the flux becomes even lower.  It is commonly maintained that most particles above the ankle have an extragalactic origin although this is by no means well-established and debate rages as to how the ankle can be explained.  There is also serious discussion of the possibility that even the highest-energy particles might have been produced in our galaxy, perhaps in association with a gamma-ray burst or a magnetar, about $10^5$ years ago [6, 7].

The interaction of a cosmic ray proton of $\sim 10^{17}$ eV with a nucleon in a stationary target nucleus produces an energy in the centre-of-mass frame equivalent to that created at the Large Hadron Collider (LHC) when protons of 7 TeV collide with 7 TeV protons.  This immediately highlights a difficulty: we are trying to interpret data in a range where key collisions between primary cosmic rays and atmospheric nuclei take place at energies well-beyond anything explored with accelerators.  Thus, when drawing conclusions from observations, it is highly desirable to avoid as far as possible assumptions about the unknown nature of the hadronic interactions at high energies.  In fact, the highest-energy cosmic rays might eventually be used to explore particle physics beyond the LHC range although this topic will not be addressed in this review.

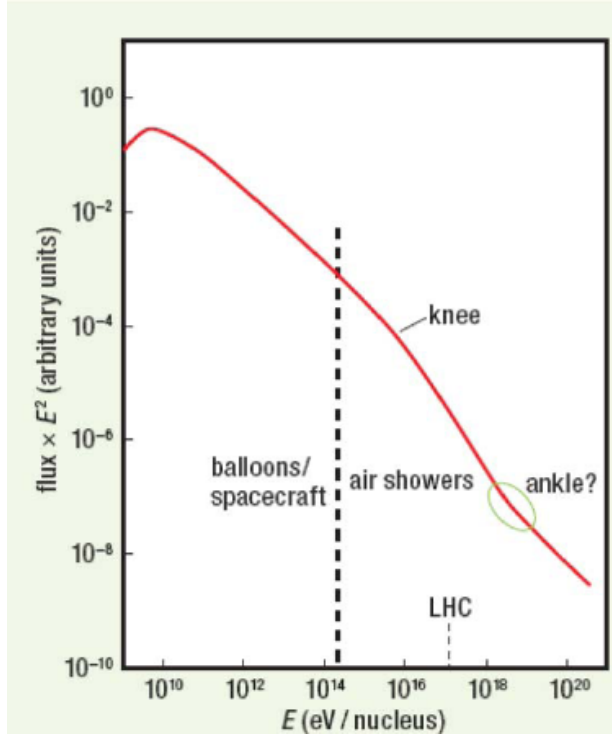

**Figure 1:** Schematic energy spectrum. The 'knee' is at ~ 3 x$10^{15}$ eV and the 'ankle' is at ~ 4 x $10^{18}$ eV. The energy at which a proton striking a stationary target would create the same centre-of-mass energy as at the LHC is indicated [4].

1.2 Why study very high-energy cosmic rays?
There are at least three reasons for studying very high-energy cosmic rays. The first is to try to answer the question: are there excesses from certain regions of the sky? Although cosmic rays were discovered over 100 years ago, we are still very uncertain as to where even the low-energy cosmic rays come from so to establish cosmic-ray astronomy would be a major breakthrough. The big problem is that between the earth and the sources there are certainly magnetic fields and the deflection of a proton in the field of our galaxy is probably about 10º even at $10^{19}$ eV. The strength and structure of the magnetic fields are also poorly known. To interpret what is seen, and perhaps to deduce the nature of the intervening magnetic fields, we really need to know the charge of the particles. This is a huge challenge as it is hard enough to determine their mass.

The second question is whether we can learn anything about the origin of the particles by measuring the spectral shape. The only firm prediction in this field is that the energy spectrum should steepen near to $4\times10^{19}$ eV if the sources are distributed uniformly throughout the Universe. This prediction was made in 1966 by Greisen [8], and independently by Zatsepin and Kuz'min [9], soon after the discovery of the cosmic microwave background radiation (CMBR) and is usually known as the GZK effect. The steepening would be expected whether the cosmic rays leaving the source are protons or heavier nuclei although computation of the details in the latter case is more complicated because of the range of nuclei and cross-sections that are involved. The input to the calculations of the shape of the spectrum requires assumptions about the nature of the mass composition at the source, the maximum energy attained at the source, the energy spectrum of the particles at production and the radiation fields through which they must propagate, to be made. Additionally it may well be that there are a range of types of source, just as there are in electromagnetic astronomy. Examples of fits to the data for different inputs will be shown in section 5.

In the rest frame of a proton of sufficient energy (typically with a Lorenz factor of ~$10^{11}$) a cosmic microwave photon of a few times $10^{-4}$ eV looks to the proton like a high-energy gamma-ray: this is the relativistic Doppler Effect. If the proton 'sees' a gamma-ray above an energy of about 150 MeV, the $\Delta^+$-resonance is produced. The $\Delta^+$ decays to a neutron and a $\pi^+$ or to a proton and a $\pi^0$: these pions and neutrons are sources of photons and neutrinos of very high energy. The proton loses about 20% of its energy in such an interaction and so the distance from which a high-energy particle can have travelled before detection is limited to ~100 Mpc. This distance is sometimes used to define 'the GZK volume' and if the suppression were to be established then one might suspect that the sources of particles above ~ $4 \times 10^{19}$ eV would lie within this volume. Of course the volume is energy-dependent as the energy loss is energy dependent. If the highest energy particles are nuclei rather than protons, the relativistic Doppler Effect is again important. Here the giant dipole resonance at ~10 – 20 MeV is relevant and nucleons, usually neutrons, are chipped from the parent nucleus one by one.

It is interesting to recall that the measurements reported in 1965 indicated that CMBR temperature was (3.5 ± 1.0) K. When I first heard of the GZK-prediction I thought that it would be possible to detect the steepening with the 12 km$^2$ air-shower array that I was then helping build at Haverah Park, near Leeds, UK. I assumed that within a few years we would see the effect. It did not work out like that: the temperature is of course 2.725 K and thus the photon density and energy threshold were changed so as to make life more difficult. Also it is now apparent that the flux of particles at high energies had generally been over-estimated in earlier work, a consequence of the difficulty of determining the primary energy. Nature often seems to arrange things so that it is as hard as possible to discover the true situation.

A third interesting question is 'how can a particle get accelerated to $10^{20}$ eV?' In a man-made synchrotron, the maximum energy reached depends on the magnetic field and radius ($E_{max}$ = ZeBR$\beta$c, where Z is the charge on the nucleus, e is the charge of the electron, B is the magnetic field in a region of radius R and $\beta$c is the velocity of the particle). The particle must be confined within the acceleration region long enough to gain energy. In a single-shot acceleration process, such as might occur near a neutron star, a high voltage could be generated between the pole and the equator: the same equation for the maximum energy holds. It is now thought to be more likely that we are dealing with diffusive shock acceleration where essentially the same relationship applies: here $E_{max}$ = kZeBR$\beta$c with the constant k < 1 and $\beta$ denoting the speed of the shock. Suitable shocks are thought to occur near active galactic nuclei, near black holes and, effective at lower energies, in association with supernova remnants. Above $10^{20}$ eV there is a dearth of objects that satisfy this equation as first pointed out by Hillas [10]. Radio galaxies, colliding galaxies, Active Galactic Nuclei (AGNs), magnetars, GRBs and perhaps galactic clusters might host the right conditions but other objects such as supernovae remnants and magnetic A stars are incapable of accelerating particles to this energy. If we could detect sufficient of these particles, then we might expect to observe a steepening in the spectrum at ~ $4 \times 10^{19}$ eV, which could be giving us information about sources out to ~70 – 200 Mpc, depending on the energy. If the particles are protons, the deflections might be small enough to reveal point sources, provided there are not too many of them.

So the problem is well-defined: we have to measure the energy spectrum to test the GZK prediction, the arrival direction distribution to explore whether we can identify sources, and the mass composition for interpretation: the snag is the tiny flux of particles in this energy range.

That the sources are likely to be rather rare can be understood from a very general argument initially due to Greisen (1965) [11] which has been confirmed and developed by Aharonian et al. [12]. Assuming that the acceleration region must be of a size to match the Larmor radius of the particle being accelerated and that the magnetic field within it must be sufficiently weak to limit synchrotron losses, it can be shown that the total magnetic energy in the source grows as $\Gamma^5$, where $\Gamma$ is the Lorentz factor of the particle. For a proton of $10^{20}$ eV this is $\gg 10^{57}$ ergs and the magnetic field must be $< 0.1$ Gauss. This analysis leads to the conclusion that putative cosmic ray sources might be strong radio emitters with radio powers $\gg 10^{41}$ ergs s$^{-1}$, unless protons or heavier nuclei are being accelerated and electrons are not. Analyses such as that in [12] show that this inequality is at least $>10^{44}$ ergs s$^{-1}$ and amongst the few nearby sources that satisfy this limit are Centaurus A and M87. It is clearly of immense interest to identify such objects to try to gain an understanding of the mechanisms by which particles can be pushed to such vast energies. To do this we need to be able to measure the energy and direction of the UHECR with good resolution.

Of course identifying a suppression of the spectrum does not automatically imply that we have seen the GZK effect. It could be that the sources become incapable of accelerating particles beyond roughly the same energy: this possibility of 'source exhaustion' will be discussed in section 6.

Interest in the field has grown in recent years and there have been several reviews. Articles that are cover the topic in more breadth than will be done here include [13, 14, 15, 16 and 17] which cover experimental data in some detail while Kotera and Olinto [18] have given an extensive overview of the astrophysical implications of the existence of such high-energy particles.

2. The Physics of the Greisen-Zatsepin-Kuz'min effect
As stated above Greisen-Zatsepin-Kuzmin (GZK) effect is a manifestation of the relativistic Doppler Effect. If a proton or nucleus with a Lorentz factor $\Gamma$ is moving through a radiation field of average energy $\varepsilon$ then the particle will 'see' a photon of energy E, where

$$E = \varepsilon\Gamma(1 + \cos \theta), \qquad (1)$$

$\theta$ being the angle between the respective directions of travel of the photon and the particle. A particularly lucid discussion of this process has been given by Stanev [19].

If the boosted-energy of the photon is above some threshold energy, $E_{th}$, then the corresponding process can take place. For illustration we consider the case of the CMBR with T = 2.725 K, average photon energy $\varepsilon = 6.4 \times 10^{-4}$ eV so that for

$$\gamma_{2.725 K} + p \rightarrow \Delta^+ \rightarrow p + \pi^o \text{ or } n + \pi^-, \qquad (2)$$

with E = 150 MeV, the threshold is $\Gamma = 1.17 \times 10^{11}$, or $\sim 1.1 \times 10^{20}$ eV for the $\Delta^+$ resonance to be excited in a head-on collision with a photon of this energy. About ~1 % of photons have an energy 3 times the mean value so calculation in the high-energy tail is very important.

The first analyses were made independently by Greisen [8] and by Zatsepin and Kuz'min [9] who set down the essential features and the importance of the process. Taking modern values, the density of CMB photons is n ~ 400 cm$^{-3}$ and the mean cross-section, $\sigma$, for the $\gamma$-p reaction is about ~ 200 µb, so that the mean free path for interaction is ~ $1.25 \times 10^{25}$ cm (or ~ 4 Mpc) and the distance scale for energy loss is L = (E/$\Delta$E) (n$\sigma$)$^{-1}$. The fraction of energy lost by the proton in each interaction is ~ 0.2 so the relevant length scale of ~ 20 Mpc corresponds to ~ $2 \times 10^{15}$ s. This is very much less than the expansion time of the Universe and thus photo-pion production is the dominant form of energy loss for high-energy protons. Of course the spectrum of blackbody radiation contains a range of photon energies so that detailed calculations are needed.

In both the Greisen and the Zatsepin and Kuz'min studies attention was drawn to the fact that the Doppler Effect would also be important for interactions between the CMBR and nuclei. Additionally Greisen pointed out that pair-production of electrons by the CMB photons would arise above ~ $10^{18}$ eV, with the fractional energy loss per interaction being ~$10^{-3}$.

Subsequent to these seminal papers, more detailed calculations were carried out by many people. Early extensions to the GZK ideas were made by Hillas [20] who examined the spectrum of cosmic rays produced in an evolving Universe, taking into account the fact that the temperature of the radiation was higher in the past. He was the first to study the pair-production effect in any detail. Stecker [21] calculated the lifetime and mean free path of protons using updated values for the temperature and for the characteristics of the interactions.

Early work on the impact of photodisintegration on nuclei heavier than protons were first explored by Stecker [22] who calculated the lifetime of helium and iron nuclei against interactions with the CMBR. Puget, Stecker and Bredekamp went on to examine photodisintegration for a variety of nuclei using detailed cross-section information and the Monte Carlo technique [23] to determine the lifetime of nuclei while Hillas [24] and Berezinsky, Grigor'eva and Zatsepin [25] made the first estimates of the energy spectra of nuclei establishing that the steepening occurs at a similar energy to that for protons.

More recently there have been further discussions about photodisintegration taking into account improvements in information relating to the cross-sections for photodisintegration and propagation in inter-galactic space [26, 27]. Nowadays it is possible to make extensive use of Monte Carlo studies of the evolution of the processes involved so that further understanding has been obtained. However, essentially no new revelations have resulted although it continues to be clear that the differences in the spectral shape for different arriving nuclei are likely to be rather small.

A particularly interesting study [27] has included the effect of inter-galactic magnetic fields. It is shown that unless the field is stronger than a few nG the main features of earlier analyses are unaltered. The strength of the magnetic field plays a role in the propagation time of the nuclei as those of greater charge will have more turbulent, and thus longer, trajectories. A striking plot derived from this work is shown in figure 2 [28]. For this calculation it has been assumed that the mass distribution of the cosmic rays injected by the sources is similar to that deduced for cosmic ray sources from direct observations of primary cosmic rays at lower energies. What is striking is that very few He or C, N, O nuclei remain even after distances as short as 20 Mpc while, by contrast, the proton and Fe survival probabilities are rather similar. This is an important result as it forces the conclusion that little can be learned about the mass of the primary cosmic rays, especially whether they are protons or Fe nuclei, by studying the shape of the energy spectrum.

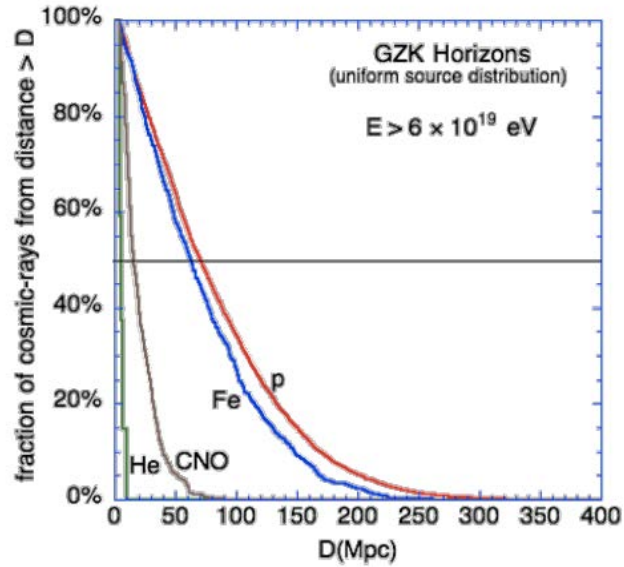

**Figure 2:** Illustration of the survival of nuclei > 6 x $10^{19}$ eV as a function of distance. Effectively only p and Fe nuclei can reach earth from sources more than ~ 50 Mpc distant [28].

There will be some discussion about what can be inferred from the spectral shape after the experimental data have been described. For now it is important to stress that the GZK effect gives no prediction about the absolute flux of particles and that the detailed spectral shape depends on the input spectra, the maximum energies and the spatial distribution of sources as well as on the composition of the material that is accelerated. It is quite possible that several different classes of object will act as sources.

3. Methods of Studying the Highest-Energy Cosmic Rays
In this section an introduction to the methods used to study extensive air showers is given with the older detector systems taken as examples. Those familiar with these techniques may wish to skip to section 4 where the more recent measurements of the energy spectrum are discussed. Throughout I avoid detailed discussions of analysis methods: appropriate references are given.

3.1. Introduction
During the 1930s the dominant motivation for investigating cosmic rays was the diverse features of particle physics that the phenomenon exposed. Inferences about the spectrum of cosmic rays arriving at earth were somewhat restricted and came mainly from analysis of variations of cosmic-ray intensities as a function of geomagnetic latitude and altitude. Energies of 1 – 10 GeV were deduced to be typical of the incoming radiation. The work of Compton, Alvarez and Johnson, building on the important input of Rossi, showed that more particles arrived from the west than from the east leading to the conclusion that the primaries are dominantly positively charged.

In 1938, the energy spectrum of cosmic rays was shown to extend at least to $10^{15}$ eV – a jump of five orders of magnitude above the energy of the low-energy cosmic rays of a few GeV that are much more abundant. The discovery of these extraordinarily energetic particles was made as a consequence of an improvement to the resolving time of coincidence circuits from milliseconds to a few microseconds by Auger and Maze [29]. The classical method of determining the resolving time of such a circuit is to measure the rate of coincidences between two counters separated by a horizontal distance much larger than the dimensions of the counters: the observed chance rate is then proportional to the resolving time of the coincidence circuit. Auger and his colleagues found that even when counters were separated by as much as 300 m the chance rate was large compared with that anticipated from the electronic improvement made. They deduced that most of the coincidences were genuine and suggested that they were caused by electromagnetic showers, the sea-level debris of energetic cosmic rays, with the charged particles in the showers firing the Geiger counters. These cascades are usually called 'extensive air showers'. The energy of the primaries was inferred using the new ideas of quantum electrodynamics which had led in the 1930s to a theory that described the development of what were called 'cascade showers' and which had been studied in great detail using cloud chambers, particularly by Blackett and Occhialini. Earlier, during his work in Eritrea on the East-West effect, Rossi had noted [30] that *'It would seem … that from time to time there arrives upon the equipment very extensive group of particles ('sciami molto estesi di corpuscoli') which produce coincidences between counters even rather distant from each other'*. Schmeiser and Bothe [31] and Kolhöster et al. [32], prompted by several theoretical and experimental indications, led successful studies to look for extensive air showers. Auger, however, does not seem to have been aware of these efforts. A more detailed account of early studies of the air shower phenomenon can be found in [33]

What was distinctive about the work of Auger and his group was that coincidences were found even when the counter separations were extended to 300 m allowing the deduction that the largest showers detected had been generated by primaries of $10^{15}$ eV [34]: the energy was estimated under the assumption that the primaries were electrons. They concluded their paper by remarking that "*it is actually impossible to imagine a single process able to give a particle such an energy. It seems much more likely that the charged particles … acquire their energy along electric fields of very great extension*". The problem of accelerating particles to the highest known energies is a persisting one and even today the single acceleration process cannot be excluded although the extension of Fermi's original mechanism in the form of diffusive shock acceleration is thought to be more likely. For seminal discussions of the essence of the acceleration requirements see Hillas [10].

An air-shower produced by a primary cosmic ray of $10^{19}$ eV contains about $10^{10}$ particles at the maximum of the resulting cascade. The particles are dominantly electrons with about 10% muons: hadronic particles form a smaller fraction. The position of the shower maximum can now be measured to within ~ 20 g cm$^{-2}$ and on average at this energy is ~ 750 g cm$^{-2}$, an altitude of about 3 km above sea-level. Coulomb scattering of the shower particles and the transverse momentum associated with hadronic interactions early in the cascade cause the particles to be spread over a wide area and a shower from a $10^{19}$ eV primary has a 'footprint' on the ground of over 15 km$^2$. All known elementary particles - and perhaps some yet to be discovered - are created in such showers but at distances > 100 m, which, will be of interest below, only photons, electron and muons are of importance.

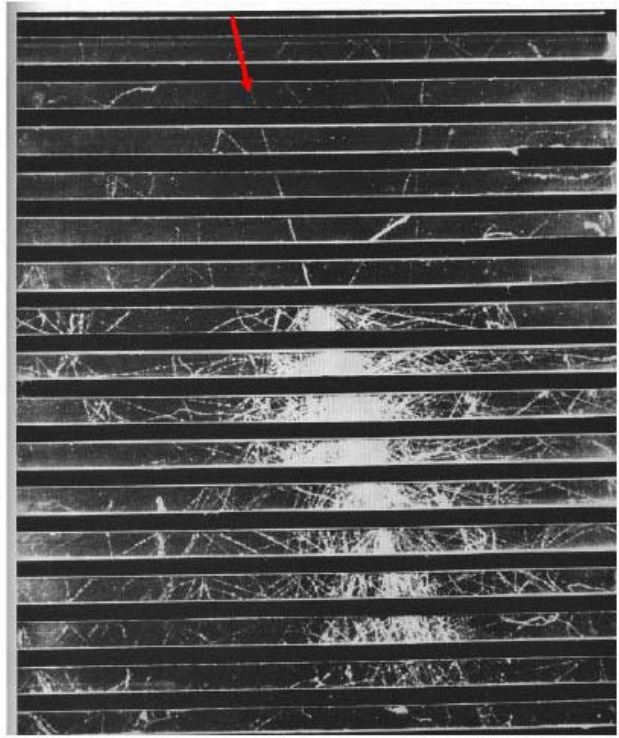

**Figure 3:** A cascade, or shower, created by a 10 GeV cosmic-ray proton developing in a cloud chamber that contains lead plates. The first interaction of the proton will most probably have been in one of the lead plates. Neutral pions feed the cascade which multiplies in the lead. Charged pions make similar interactions to protons in the lead or decay into muons. The cross-sectional area of the cloud chamber was 0.5 m x 0.3 m [35].

A picture of a miniature cascade shower, produced in a succession of lead plates inside a cloud chamber, is shown in figure 3 [35]: the energy of the proton that created the shower was probably ~ 10 GeV. The shower grows because of bremsstrahlung and pair production processes occurring in the lead with a small number of hadronic interactions providing the neutral pions that feed the electromagnetic cascade. The gas in the chamber serves mainly to make visible the growth of the cascade. All of the important features of shower development, such as the rise and fall of the particle numbers and the lateral spreading, are evident, as are muons that penetrate much more deeply into the chamber than most of the electrons. One could imagine positioning nano-detectors of particles on the lead plates and using these to measure the time of arrival of particles at the detectors as well as the number of particles striking them. The relative arrival times could be used to obtain the direction of the incoming particle. The extensive air-showers that I will be concerned with are scaled up versions of what is seen in figure 3. The showers can be visualised as giant discs of particles sweeping through the atmosphere at the velocity of light with the radii of the discs being a few kilometres and the thicknesses, depending on distance, being a few hundred metres.

3.2. Detection of extensive air showers using surface arrays and the estimation of the primary energy
The generic method for the detection of air showers is based on developments of the discovery technique used by Auger and his colleagues. The procedure is to spread a number of particle detectors (plastic scintillators and water-Cherenkov detectors have been used primarily since the mid-1950s) in a more or less regular array over the ground. The higher the primary energy the greater is the spacing that can be tolerated between the detectors and for the study of showers produced by primaries above $10^{19}$ eV the separation can be over 1 km. At such distances the relative contributions to the signal from muons, electrons and photons depend upon the type of detector and the angle of arrival of the shower with respect to the vertical. For example, water-Cherenkov detectors are typically about 1 m deep and so most of the electrons and photons, which have a mean energy of ~ 10 MeV at > 100 m from the shower axis, are completely absorbed. By contrast, scintillation detectors are usually only a few centimeters thick and the response to muons and electrons are nearly identical while photon absorption is small. The detector signal is expressed in units of the signal produced by a muon traversing the detector vertically (the vertical equivalent muon or VEM) and, at sea-level, there are ~ 5 VEM m$^{-2}$ in a water-Cherenkov detector at 1 km from the centre of a shower produced by a primary of $10^{19}$ eV: thus with a detector of 10 m$^2$ a strong response is detectable. A smaller spacing between detectors is often desirable but is usually not feasible for financial reasons except on small areas of a larger array: land-access can also be an issue. The incoming directions of the events can easily be measured to better than 2° by determining the relative times of arrival of the shower disc at the detectors. The pattern of particle densities at the ground is used to estimate an appropriate quantity that can be related to the primary energy.

In figure 4 the famous event claimed to have been initiated by a cosmic ray of $10^{20}$ eV recorded by Linsley at the Volcano Ranch array, the first of the giant shower arrays [2], is shown. The layout of Volcano Ranch is typical of a ground array and will serve to epitomise what was done subsequently. At Volcano Ranch 19 3.3 m$^2$ plastic scintillators were distributed on a triangular grid with 884 m separation with a 20$^{th}$ detector shielded with lead to give a measurement of the muon content for $E_\mu$ > 220 MeV. Specific details of other surface arrays can be found in the references given below or in the review paper of Nagano and Watson [13].

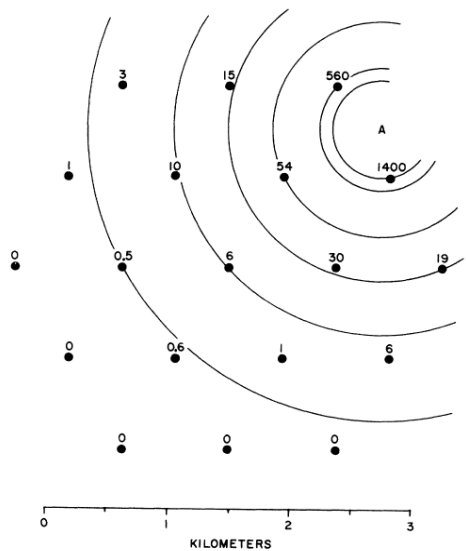

**Figure 4:** The largest event recorded at Volcano Ranch estimated to have been produced by a primary of $10^{20}$ eV [2].

Finding the energy of the particle that has created a particular shower is not straight forward and it is instructive to appreciate the approaches that have been adopted over the years. Although it was discovered relatively early that air showers contained nucleons, pions and muons in addition to an abundance of electrons and photons, the gross features of showers were found to be relatively well-described under the assumption that the primaries were electrons. It thus became the practice to infer the primary energy from a measurement of the total number of charged particles, N, – dominantly electrons and positrons – in a shower, relating this to the primary energy using theory provided by such as the Nishimura-Kamata equations [36] that describe showers produced by photons or electrons. The number of particles was straight forward to measure when the detectors were Geiger counters as these respond predominantly to charged particles. Also, for the study of the showers produced by primaries of energy less than ~$10^{17}$ eV, it was practical and economically feasible to build arrays in which the average separation of the detectors was less than the Molière radius, about 75 m at sea-level: roughly 50% of the charged particles of a shower lie within this distance. Once it became understood that there were significant fluctuations in shower development so that showers observed at an observation depth with a size N could arise from primaries of a range of energies other methods of determining the primary energy were gradually developed. It is evident from figure 3 that one cannot know from the observations at one single depth where the first interaction has taken place (and it is obviously a challenge to find the atomic mass of the primary particle). It is also clear that the determination of the energy from measurements below a single lead plate deep in the chamber, or from measurements at a single atmospheric depth, cannot be straight forward.

As better understanding of showers was acquired, there were moves away from using the photon/electron approximation to estimate the primary energy. A difficulty in finding N was recognised as scintillation counters were increasingly introduced into shower arrays towards the end of the 1950s. However, because of the success and simplicity of the approach with Geiger counters and the lack of other methods to find the energy on an event-by-event basis, considerable effort was initially expended in relating the scintillator measurements to what would have been the particle count had a Geiger counter been located at the same position. This adjustment to particle number was reasonable while the spacing between detectors remained small. For example at the Agassiz array [1] measurements were made at distances much closer to the shower core than one Molière radius and the scintillator response was converted to particle number using an array of Geiger counters operated for that purpose. The importance of the thickness of the scintillators was soon appreciated and also that the conversion factor from scintillator signal to number of charged particles depended on the distance of the scintillators from the shower core because the energy spectrum of electrons and photons is distance-dependent.

It had long been recognised that most of the energy of the primary particle was dissipated in ionisation of the air above the detectors and that by determining the '*track-length integral*' it would be possible to measure the shower energy using only experimental data with some allowance being made for energy carried by non-ionising particles such as neutrinos and for the energy carried into the ground. The particle track-length integral is defined as

$$\int_0^\infty N(t)dt, \qquad (3)$$

where N is the number of particles at a depth t. If t is expressed in radiation lengths then the energy of the primary is given by multiplying the value of the integral by $\varepsilon_c$, where $\varepsilon_c$ is the critical energy (84.2 MeV for air). It is assumed that $\varepsilon_c$ is an accurate measure of the average energy lost per radiation length by the charged particles of the shower. A general idea of the form of N(t) can be gained by inspection of figure 3 but, while in the case of the cloud chamber event the integral can be evaluated relatively directly, Greisen's first attempt [37] was made using a variety of measurements of the rate of detection of showers at sea-level, at mountain altitudes and in airplanes to obtain the shape of the longitudinal development curve. Combining measurements made with different instrumentation by many people was difficult and was confined to showers which had ~ $10^5$ particles at sea-level.

A superior method was developed by Greisen [38] using the relationship between the observed number particles and the primary energy building particularly on the studies of Cherenkov light by Chudakov and his collaborators [39] in the Pamirs. Greisen also made use of direct measurements of the energy and number of muons in showers and of the electromagnetic energy flow. Chudakov and his group found that in a shower of $1.4 \times 10^6$ particles, there were $1.2 \times 10^5$ photons from Cherenkov radiation. Assuming that the bulk of this radiation was from electrons > 50 MeV, Greisen estimated that the total ionisation loss, above the observation level, for a shower of this size was $5.2 \times 10^{15}$ eV or 3.7 GeV per particle. To get the total energy, 0.2 GeV per electron has to be added for each of the electromagnetic and nuclear particle components with a further 0.4 GeV per particle for the muons and neutrinos giving the primary energy as $6.3 \times 10^{15}$ eV.

While it is clearly not feasible to make such precise energy analyses for the components of much larger showers in which measurements close to the shower axis are impractical, the track-length integral method is the essence of the fluorescence technique (see below) which Greisen did much to promote and, through measurements of the Cherenkov radiation, remains a key technique in the energy determinations at the Yakutsk array (see below).

Returning to the experimental observations, it was the shower-array technique that was first used to claim that the cosmic ray spectrum extends beyond $10^{19}$ eV using the array at Volcano Ranch (figure 4) [2]. The large spacing between the detectors presented difficulties for determining the number of particles in the shower. At the atmospheric depth of Volcano Ranch, 834 g cm$^{-2}$, the Molière radius is ~ 100 m and for two very large events of energies > $10^{19}$ eV [1, 2] it was argued that the size, N, could be inferred by integrating an average lateral distribution over all distances with the energy being deduced under the assumption that the showers were at the maximum of their development at the depth of observation. The conversion to primary energy was based on Greisen's method of the track-length integral. It is now known from direct measurements that the average depth of maximum, $X_{max}$, is ~ 100 g cm$^{-2}$ higher in the atmosphere than was then believed so that some of the inferred energies made by the MIT and Volcano Ranch groups using this method were perhaps too low. The estimates of energies at various arrays have been re-assessed over the years as our understanding of the air-shower phenomenon has increased but Linsley's initial evaluation of his largest event [2] has proved to be rather robust.

The distributed detector concept was further developed by a group from the University of Leeds who constructed an array at Haverah Park which differed in a number of respects from that at Volcano Ranch [40]. The detector elements were deep (1.2m) water-Cherenkov detectors of large area (typically 13.5 and 34 m²) and were spread over 12 km$^2$. The large areas of the peripheral detectors enabled signals to be measured as far as 3 km from the axis. Water-Cherenkov detectors respond very efficiently to photons and electrons in addition to muons and, at 1 km, the total response comes from ~40% muons and ~60% electrons and photons in vertical showers of ~$10^{18}$ eV. At Haverah Park the tanks were 3.4 radiation lengths thick so most of the electromagnetic component was completely absorbed.

The determination of the energy of the primary presented particular problems as it was impossible to relate the observed signal to the number of charged particles in a reliable way. Measurements made with muon, scintillator and water-Cherenkov detectors at the same point in the shower, showed that on average for every electron there were roughly 10 photons [41] and that the mean energy of both components was about 10 MeV although these numbers changed with distance from the shower core. The first efforts to go directly from what was observed to the primary energy was to measure the energy deposited in a pool of water envisaged as an annulus of inner and outer radii of 100 and 1000 m and with a depth of 1.2 m. This deposit, $E_{100,}$ was related to primary energy using model calculations developed by Hillas and Baxter in the late 1960s. Empirical descriptions of the hadronic features of showers were combined with calculations of the electromagnetic cascades in early Monte Carlo calculations. At 500 m from the axis of a vertical event produced by a primary particle of ~ $10^{18}$ eV about half of the energy flow is carried by electrons and photons of mean energy ~ 10 MeV that are very largely absorbed in the water with the remaining energy being carried by muons. It was found the $E_{100}$ was ~ 1/160 of the primary energy [42] and this method of estimating the primary energy was used to argue that a shower of energy > 5 x $10^{19}$ eV had been detected at Haverah Park soon after the Greisen-Zatsepin-Kuzmin prediction that such events should be rare [43].

Two problems with this approach were soon identified: firstly, in large events there was rarely a detector close to 100 m from the shower axis and secondly, lack of knowledge of the variation of the lateral distribution function with energy led to a systematic uncertainty that was hard to evaluate. The problem of what parameter to measure was solved in 1969 through a seminal insight by Hillas [44] and his solution has been widely adopted in subsequent measurements with the ground arrays at AGASA, the Auger Observatory and the Telescope Array. Hillas, using a sample of 50 events recorded at Haverah Park, showed that if $E_{100}$ was found using different power laws to describe the lateral distribution function the differences in the derived values was large, ~ 1.7. It is worth quoting Hillas directly from his paper

*"However, because of the geometry of the array, the alternative fits to the data obtained with the different structure functions* [now called lateral distribution functions] *are found usually to give the same answer for the density $\rho_{500}$ at 500 m from the axis: in the sample examined $\rho_{500}$ was usually altered by less than 12% by the different assumptions. A similar effect will arise in other large arrays, the exact distance R for which the density is well-determined depending on the detector spacing."*

For the larger Haverah Park array, which was being brought into operation at the time of Hillas's work, the parameter $\rho_{600}$ was adopted with S(600), S(800) and S(1000) subsequently chosen for AGASA and the Telescope Array scintillator systems and the Auger water-Cherenkov array respectively: S(r) denotes the signal at the appropriate distance.

The merit of such a parameter can be seen in figure 5 where three very different lateral distribution functions are shown. For any shower and for a wide but physically sensible range of lateral distribution functions, there is a crossing point at which the spread in the signal size is a minimum. The distance of this crossing point has been named $r_{opt}$, the optimal distance for a size parameter associated with that event. For a particular array spacing there will be an average $r_{opt}$ for a set of showers as has been discussed in detail [45]. For $\rho_{500}$, which was replaced by $\rho_{600}$ when the average spacing between the detectors at Haverah Park was increased, Hillas and colleagues [46] showed that the energy deduced in the $10^{18}$ eV to $10^{19}$ eV range was independent of mass and models at the 10 – 15% level. The relationship between the energy and $\rho_{600}$ is given by $E = \beta \rho_{600}^{\alpha}$ where α and β are found from models. For all models explored in [46] that fitted a wide range of experimental data available at the time, α ~ 1 with values of β having a scatter of ~ 20% around 7 x $10^{17}$ eV and the scatter of α around the preferred value of 1.18 being around 3%.

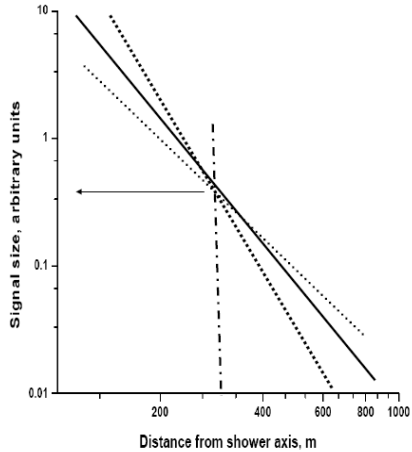

**Figure 5:** Illustration of how different lateral distributions allow a measurement, $S(r_{opt})$, that has a small spread. The lateral distributions are adequately represented by power-laws for this purpose though the differences in slope are exaggerated in the sketch. An appropriate choice of distance, $r_{opt}$, is available that depends only on the spacing of the shower detectors [45]. In this example $r_{opt}$ is ~ 350 m.

Prior to the development of the Pierre Auger Observatory, to determine the primary energy from measurements with a surface array one made use of predictions from calculations of shower development that drew on extrapolations of the properties of interactions studied at accelerators. Returning to figure 3, the challenge is to measure the energy of the incoming particle using the information gleaned from a set of detectors at one level in the atmosphere. Fluctuations are a nightmare: one cannot know from observations at a single level where the first interaction has taken place and it is clearly difficult to identify the atomic mass of the primary particle. Furthermore the characteristics of hadronic interactions are uncertain at centre-of-mass energies of $\sqrt{s} \sim 10^2$ TeV that are of interest at $10^{19}$ eV. Recall that the LHC is designed to operate *only* at $\sqrt{s} \sim 14$ TeV. Additionally, around 50% of the particles in the shower lie within about 100 m of the shower core (the Molière unit in air at sea level is about 75 m) and so are usually unobservable with an array of sparsely spaced detectors. Also the systematic uncertainty is hard to quantify (one is almost tempted to say *'impossible'* because of the unknown hadronic physics).

While the electromagnetic processes that produce most of the particles in the shower are well-understood that is not the case for the hadronic interactions which produce the neutral pions so a word is appropriate here about the models of these interactions that are used to interpret air-shower observations. The particle that is most critical to the development of the cascade leaves the collision between an incoming proton and a nucleus, with high energy and small transverse momentum. This is called *the leading particle* (usually a proton or a neutron) and on average carries ~ 0.5 of the incoming energy. The leading particle could be a pion which would typically carry ~ 0.2 of the primary energy. Also important are the associated high-energy pions. These particles have experienced low momentum transfer and cannot be modelled by QCD. In the early days of Monte Carlo calculations (1960s to 1980s) the emphasis was on the use of phenomenological models such as those used by Hillas et al. [46] where cross-sections, multiplicity, transverse momentum and inelasticity were inserted largely empirically.

These models were gradually superseded by others that are more soundly based theoretically and use has been made of the growing store of information from accelerator measurements at higher and higher energies. The models of hadronic physics that are currently available to interpret air-shower data and will be encountered in the discussions that follow and in many of the papers cited in this review are from the QGSJET and of SIBYLL families. The QGSJET set is based on the Gribov-Regge theory [47] which successfully describes elastic scattering and cross-sections. The Sibyll group is more phenomenological and was based originally on the mini-jet model although recently additions have been made to give a more accurate treatment of soft collisions. Accessible accounts of the different models can be found in [48, 49, 50]. Models are tuned to the latest data from accelerators (increasingly the LHC data are being used) and then extrapolations are made to $10^{20}$ eV. A clear account of these extrapolations has been given in [51] where there is also a lucid explanation of the differences between the SIBYLL and QGSJET models. It is worth remembering in what follows that the QGSJET models predict a slower rise of cross-section with energy than SIBYLL: at $10^{20}$ eV the difference is about 100 mb. There are similar differences between the cross-sections predicted for π-air collisions. There are also large differences between the multiplicities predicted by these two models with the QGSJET values being roughly twice as high as those from SIBYLL. The same trends are associated with π-air collisions. The multiplicity is important in determining the number of muons expected in the shower. Recently the EPOS model [52] has been developed in which the production of nucleon-anti-nucleon pairs is included. This leads to enhanced muon production by comparison with QGSJET.

A simple and elegant way of understanding the development of extensive air-showers has recently been set out by Matthews [53]. A study of his approach is strongly recommended before trying to understand the more detailed simulations which are often carried out using the CORSIKA Monte Carlo framework developed by the KASCADE group [54].

A group from the University of Sydney [55] constructed an array of buried liquid scintillation detectors at Narribri covering an area of ~ 100 km$^2$. Some highly original ideas were incorporated within this device, several of which were used in the design of surface-detector array of the Pierre Auger Observatory. The detectors were operated autonomously and coincidences between stations determined off-line (often many days later) by matching triggering times derived from a signal beamed across the array from a transmitter. The arrival directions were found in the standard manner. Only muons were detectable making the energy determination unusually dependent upon the choice of model. Furthermore the data relating to shower size were compromised by a serious problem with after-pulsing in the photomultipliers and for these reasons finding reliable values of the energy was difficult. The Sydney spectrum is not usually shown in summaries.

3.3. Detection of Extensive Air Showers using Fluorescence Radiation
Fortunately there is another detection method which does not suffer greatly from the reliance on hadronic interaction models to determine the primary energy that plagues surface arrays, although it presents other challenges. A charged particle travelling through the atmosphere can excite the 1N and 2P bands of nitrogen, with the major component of the resulting fluorescence emitted in the 300 - 450 nm range. Although only about 4.5 photons per metre of electron track are radiated, the light is visible with the naked eye, as are aurorae, when very many particles are incident on the atmosphere simultaneously. In the 1960s [56] there was interest at Los Alamos in using fluorescence radiation to monitor explosions of nuclear bombs while at about the same time the idea of using the earth's atmosphere as a calorimeter to detect ultra-high energy cosmic rays occurred to several people; Suga in Japan [57], Chudakov in Russia [58], and Greisen in the USA [11]. The fluorescence light is radiated isotropically thus offering the potential of observing an air shower far from the direction of travel of the bulk of the particles in the shower along the direction of the incoming primary particle. In principle this technique allows the track-length integral (equation 3) to be measured directly.

Interesting histories of the early days of fluorescence studies have been given by Linsley [59] and Tanahashi [60]. It seems certain that the first detections were by Suga's group [61] at Mt Dodaira near

Tokyo (see also the discussion of Dawson [62]). A more extended account of the history of the development of this technique can be found in [33]. Neither the Japanese site nor the location at Cornell were ideal for observing the very faint fluorescence light and significant progress was made only when the idea was championed at the University of Utah by a group led successively by Keuffel, Bergesson and Cassiday. The instrument developed there was known as Fly's Eye. An important first step was the coincident observations of 1977 between showers detected with the scintillator array at Volcano Ranch and with a 3-telescope prototype of the Fly's Eye detector [63]. This effort yielded the first unambiguous demonstration of the detection of fluorescence radiation produced by air showers.

After the work at Volcano Ranch the Fly's Eye system was operated at the Dugway Proving Grounds in Utah (40º N , 860 g cm$^{-2}$) for many years and it is worth describing this ground-breaking instrument and the method of analysis in some detail [64]. Two 'eyes' were developed: the first consisted of 67 spherical mirrors each of 1.5 m diameter and with 12 or 14 photomultipliers at their focus. The mirrors were arranged so that whole sky could be imaged, with each photomultiplier able to view a hexagonal sky-region 5.5º in diameter. A second eye, 3.4 km from the first, contained 36 mirrors and viewed the half of the sky in the direction of the first set of telescopes. There were 880 and 464 photomultipliers in the first and second telescopes respectively.

A detailed description of the determination of the shower direction is given in [64] and only an outline is given here. The classic diagram showing the geometry needed for reconstruction, taken from Bunner's Doctoral Thesis, is in figure 6 [65].

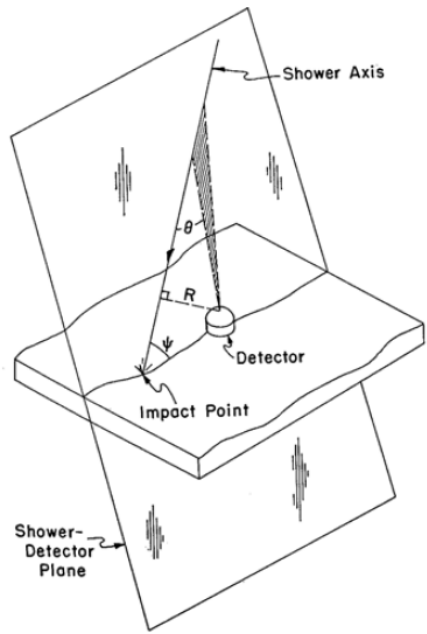

**Figure 6:** Perspective view of the shower geometry used for reconstruction of fluorescence events as created by Bunner [65].

The first step is to fit a plane to the pattern of photomultipliers that have signals above the night-sky background. The incident angle $\psi$ in this plane and the impact parameter R from the fluorescence detector define the trajectory. These parameters, and the times at which the photomultiplier tubes trigger, are related by:

$$t_i = t_o + (R/c) \tan(\theta_i/2), \qquad (4)$$

where c is the velocity of light, $t_i$ is the time of trigger of the i-th tube and $\theta_i$ is related to $\psi$ by

$$\theta_i = \pi - \psi - \chi_i, \quad (5)$$

with $\chi_i$ being the elevation angle of the photomultiplier and $t_o$, the time at which the shower plane passes through the detector. For short tracks, it is difficult to distinguish the function of equation (4) from a straight line and R and $\psi$ are not identified independently. For good direction determination with a single eye, a track length of at least 40° is required. However, if a shower is seen simultaneously by both eyes of a stereo system, the intersection of the planes determined with each eye gives a superior determination of the shower direction. With the stereo technique, the angular accuracy can be as good as 0.6°.

An understanding of the atmospheric conditions between the track of the shower and the fluorescence detectors is very important when determining the number of charged particles within a particular element of the longitudinal profile as in extreme cases the light may have come from 50 km away. Moreover, unlike a calorimeter at an accelerator, the medium is dynamic. The dependence of the fluorescence yield must be known accurately as a function of temperature, pressure and humidity at the altitude at which the light is emitted. When estimating the transmission through the atmosphere, Rayleigh and aerosol scattering must be taken into account and accordingly the atmosphere must be well-understood and be regularly monitored.

A difficult background is Cherenkov radiation produced close to the direction of travel of the shower particles and scattered by molecules and aerosols: the Cherenkov light, ~ 30 photons per metre of electron track, is emitted predominantly in the forward direction at ~ 1° from the track of a particle, and thus the light signal builds up as the shower develops in the atmosphere. Thus, in the worst case, if a shower is moving towards a fluorescence telescope, the Cherenkov radiation will dominate in the light received. Scattered Cherenkov light from the lower regions of the atmosphere is a serious problem demanding careful atmospheric monitoring.

The number of charged particles in the shower in each angular bin defined by the aperture of a photomultiplier is determined iteratively from the number of photoelectrons detected at photomultipliers after corrections have been made for scattered Cherenkov radiation and the attenuation between the detector and the shower caused by aerosol and Rayleigh scattering [64]. The resultant profile is fitted either with a 3-parameter function due to Gaisser and Hillas [66] or with a Gaussian function. The Gaisser-Hillas function is

$$N(X) = N_m \left( \frac{X - x_0}{X_{max} - x_0} \right)^{(X_{max} - x_0)/\lambda} e^{(X_{max} - X)/\lambda}, \quad (6)$$

where $x_0$ is the depth of the first interaction and $\lambda = 70$ g cm$^{-2}$. $X_{max}$ is the depth of shower maximum and X is the atmospheric depth.

The cascade profile gives the particle number as a function of depth and the energy of the primary can be obtained rather directly. Essentially this is the track-length integral approach developed by Greisen [37, 38] and encapsulated in equation (3). The energy of the particle that initiates each cascade is found by integrating under the cascade curve and multiplying the result by the ratio of the critical energy for electrons in air ($\varepsilon_0 \sim 84$ MeV) to the radiation length ($X_0 \sim 37$ g cm$^{-2}$). Thus the energy is given by:

$$E = (\varepsilon_0/X_0) \int N_e(x) \, dx = (2.27 \text{ MeV}) \int N_e(x) \, dx, \quad (7)$$

where the integral is the area under the curve fitted to the measurements. An energy-dependent correction of about 10%, that will be discussed in more detail below, must be made for 'missing energy' that goes into muon, neutrino and hadronic channels which travel into the ground. In addition assumptions, based on a variety of measurements, as to the efficiency with which the energy in the shower is converted into fluorescence light, the fluorescence yield, must be made will also be discussed later.

With the stereo system of Fly's Eye an energy spectrum was obtained up to $4 \times 10^{19}$ eV, too low to assess the GZK prediction in detail [67]. In addition, and most remarkably, an event claimed to be of $3 \times 10^{20}$ eV was recorded [68]. The longitudinal development of this shower is shown in figure 7, and is intended as an iconic point of reference like the cloud chamber event of figure 3.

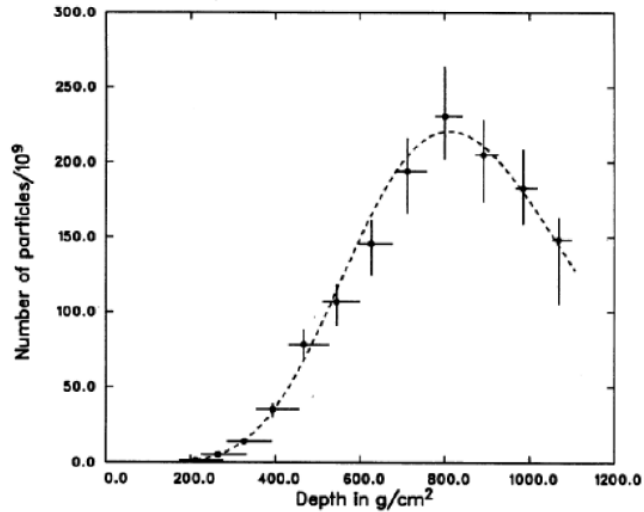

**Figure 7:** The event recorded by the Fly's Eye group with an energy estimated as $3 \times 10^{20}$ eV [68]. The rise and fall of the cascade is very clear as is the position of shower maximum at ~800 g cm$^{-2}$. This parameter is of great importance for estimating primary mass.

The outstanding pioneering efforts of the Fly's Collaboration began in the 1970s and led to the construction of an improved instrument (HiRes) operated by a US consortium, again at the Dugway site, and to the inclusion of fluorescence detectors in the Pierre Auger Observatory and the Telescope Array. Not surprisingly, as well as instrumental improvements in these second and third generation detectors, there have been refinements in the methods of analysis. Where appropriate, these are discussed in section 4.

3.4. Detection of Extensive Air Showers using Cherenkov Radiation
As mentioned above, Cherenkov radiation is produced along with fluorescence radiation by the particles of the shower as it develops in the air. The detection of this radiation has been exploited particularly for the study of showers produced by primaries of energy $< 10^{18}$ eV and a useful survey of the technique can be found in [69]. Because the Cherenkov radiation is emitted in the forward direction, the lateral spread of the light is similar to that of charged particles. This feature, coupled with the small on-time available for any optical technique, means that Cherenkov radiation is not seen as being particularly useful for monitoring the vast areas needed to study the highest energy events. However, the light signal is strong and the total signal, integrated over the shower front, is closely proportional to the ionisation lost by charged particles in the atmosphere. Hence measurement of the Cherenkov radiation provides another route for measuring the energy of the primary. It is also argued that the depth of shower maximum can be extracted from the lateral distribution and the temporal distribution of the signal, though recourse to model calculations is needed to achieve this. This technique has been used at Yakutsk in Siberia for the study of the highest energy cosmic rays: the Yakutsk array is singular in this regard at the highest energies [75].

## 4. The energy spectra reported from the AGASA, HiRes, Yakutsk, Auger and Telescope Array Observatories

Results reported from the early instruments (Volcano Ranch, Haverah Park, Narribri and Fly's Eye) have been overtaken in the statistical sense by the much greater exposures accrued with HiRes, AGASA, Yakutsk, the Pierre Auger Observatory and the Telescope Array although some individual events of high-energy recorded by the early devices remain of interest. Despite having a rather small collecting area, an array is still operational at Yakutsk that has some unique features. In this section brief descriptions of the instrumentation at AGASA, Yakutsk, HiRes the Pierre Auger Observatory and the Telescope Array are given, along with the measurements of the energy spectrum reported by the different groups. A critical discussion is deferred for section 5.

4.1. Energy Spectrum from the AGASA Observatory

The Akeno Giant Air Shower Array (AGASA), operated at a mean atmospheric depth of 920 g cm$^{-2}$ and a latitude of 35º 47' N by a collaboration led by the Institute of Cosmic Ray Research of the University of Tokyo, contained 111 plastic scintillation detectors, each of 2.2 m$^2$, laid out on a grid of about 1 km spacing. The muon detectors installed near 27 of the 111 stations are not relevant to the present discussion. The scintillators were 5 cm thick and so responded mainly to electrons and muons. The detectors were connected and controlled using a sophisticated optical-fibre network. A total of 11 events with energies above $10^{20}$ eV have been reported with two being above $2 \times 10^{20}$ eV, one of which is described in detail in [70]. When the AGASA instrument was switched off in January 2004 the majority of events claimed as being from primaries of $> 10^{20}$ eV had been reported from it and this work on its own appeared to provide convincing evidence for trans-GZK particles [71]. The exposure achieved was ~ 1770 km$^2$ sr yr.

For the AGASA studies, the shower size parameter chosen was the scintillator signal at 600 m from the shower axis, S(600), where the signal size, S, at a detector is measured in units of the Vertical Equivalent Muon (VEM). The thinking behind this approach was guided by the Hillas work discussed in section 3.2 [44]. The relationship between S(600) and the primary energy must be found using Monte Carlo simulations of the shower development. Typically S(600) can be measured with an accuracy of ~10%. The observed value of S(600) for a given energy depends on the angle of incidence of the shower and a correction has to be made for this effect as shower simulations are sometimes available for only a restricted range of angles of incidence. It is assumed that the flux of cosmic rays is sufficiently isotropic to allow the inference that, if the integral rate of events above a particular size arriving from different zenith angles is identical, then the primary energy corresponding to that size is the same. This is known as the 'constant intensity method' (CIC) and has served the ground-array community well for over 40 years [72]. For AGASA it was found empirically that, if the depth of observation is greater than the depth at which the showers reach maximum, the attenuation of the size, S(600), is close to exponential. In the case of AGASA, the reference angle is taken as 0º and the relationship between $S_\theta(600)$ and $S_0(600)$ is

$$S_\theta(600) = S_0(600) \exp\left[-\tfrac{X_0}{\Lambda_1}(\sec\theta - 1) - \tfrac{X_0}{\Lambda_2}(\sec\theta - 1)^2\right], \tag{8}$$

where $X_0 = 920$ g cm$^{-2}$, (the vertical atmospheric depth at AGASA), $\Lambda_1 = 500$ g cm$^{-2}$ and $\Lambda_2 = 594\,^{+268}_{-120}$ g cm$^{-2}$ (no uncertainties are given for $\Lambda_1$) [73] . $S_0(600)$ is an estimate of the S(600) that an event arriving at an angle θ would have had had it arrived in a vertical direction. The uncertainty in the determination of $S_0(600)$ due to the uncertainty of the attenuation curve is ~ 10% [73].

The conversion from $S_0(600)$ to primary energy in the case of AGASA was explored with a variety of models and mass compositions [71] with the hadronic interaction model adopted as reference being QCDJET, a model similar to the QGSJET model discussed above [74]: the primary particles were assumed to be protons. The relationship derived is of the form $E = a\, S_0(600)^b$, where a and b are found from the simulations. For the average atmospheric depth across the AGASA site the relation adopted is

$$E = 2.13 \times 10^{17}\, S_0(600)^{1.0}\ \text{eV}. \tag{9}$$

For all of the models used, if iron nuclei are assumed rather than protons, then the estimate of the primary energy is reduced, but by less than 5%. The energy spectrum derived from these procedures is shown in figure 8.

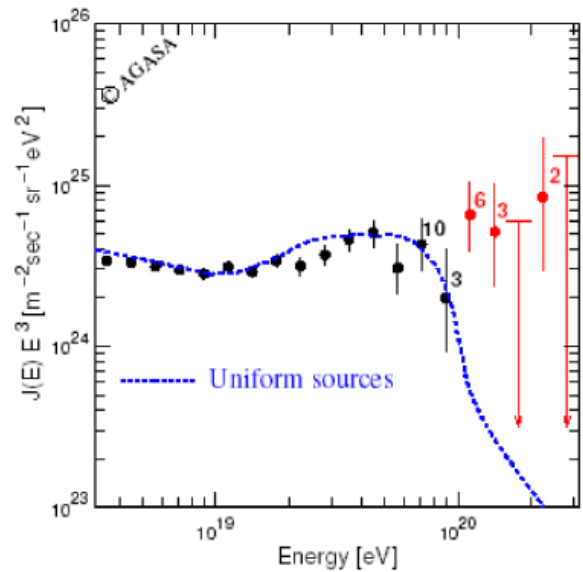

**Figure 8:** Energy spectrum reported by the AGASA group [71].

4.2. Energy Spectrum from the Yakutsk Shower array
Like the AGASA Observatory, the installation at Yakutsk uses plastic scintillation counters as the workhorse detectors. The array currently covers 10 km$^2$. It is situated at 100 m above sea-level (1020 g cm$^{-2}$) and at latitude 61° 36' N and was earlier operated for many years with the detectors covering 25 km$^2$. Presently it consists of 58 scintillation counters on the surface and 6 underground detectors that are used to measure muons. Since its inception in 1970, a unique feature has been the measurement of air-Cherenkov radiation and 48 vertically-mounted, 15 cm diameter, photomultipliers are used for this work. The exceptionally clear Siberian skies make the data from the Cherenkov detectors particularly useful and allow a different, less model-dependent, approach to measurement of the cosmic ray spectrum from that of the AGASA group. During the winter observation periods the temperature is – 40 C and absorption of Cherenkov light in the atmosphere is negligible: only molecular and aerosol scattering are considered. Details of the monitoring process can be found in [75]. The Cherenkov light measurements are used to find the energy lost by electrons through ionisation, $E_i$, essentially using the concept of the track-length integral method discussed above (section 3.2 and equation 3). At $10^{19}$ eV, $E_i$ accounts for about 77% of the energy of the incoming cosmic ray and is estimated from the total Cherenkov light, $Q_{tot}$, reaching the ground by integrating over all distances using an empirically-measured lateral distribution functions. The relationship deduced is

$$E_i/Q_{tot} = (3.01 \pm 0.36) \times 10^4 (1 - X_{max}/(1700 \pm 270)). \qquad (10)$$

The estimate of $E_i$ requires knowledge of $X_{max}$ (in g cm$^{-2}$) which is derived from the lateral distribution of Cherenkov light [75]. Estimates of the ionisation loss of electrons travelling into the ground (~15%), the energy in the muon component (4%) and the energy which is unobserved (neutrinos and high energy muons, ~4%) must be added to the more direct measurement of $E_i$. The estimate by the Yakutsk group of the uncertainties in the derivation of the energy is 32% and this figure has been used to unfold the experimental uncertainties and so deduce the true energy spectrum.

The spectrum thus derived is shown in figure 9. The spectrum deduced from events in which there are Cherenkov light measurements is in excellent agreement with results at lower energies also made with arrays of Cherenkov light detectors [75]. The ankle feature is seen clearly but the method is limited at the highest energies by lack of exposure. Moreover a previous spectrum from the Yakutsk group based on the scintillation detectors is in poor agreement at the lower energies and the differences are not yet understood within the Yakutsk group.

The reader may wish to note that 4 events are claimed with energies above $10^{20}$ eV where the size has been deduced from measurements of S(600), the latter being calibrated using the Cherenkov light measurements to estimate the energy. However, no flux estimate has been reported for these events which were not discussed in [75], the definitive report from the Yakutsk group. Therefore for the purposes of this review only the energy spectrum based on the Cherenkov light work (figure 9) will be considered.

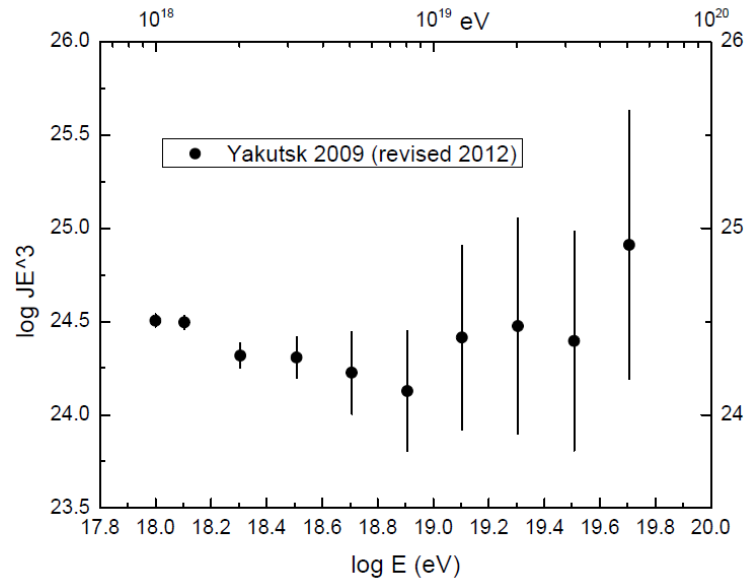

**Figure 9:** The energy spectrum reported by the Yakutsk group based on Cherenkov light measurements [75] revised according to adjustments report by A Ivanov (private communication, March 2012).

4.3. Energy Spectrum measured with the HiRes Detectors

A stereo system of fluorescence detectors known as HiRes (the High Resolution Fly's Eye) was developed by a US team at the Dugway site as a second-generation instrument. Compared with the Fly's Eye device, the two telescope systems of HiRes were placed 12.6 km apart (as against 3.4 km formerly). The telescopes operated from 1997 – 2006. One of the detectors (HiRes-I) consisted of 21 telescopes that together covered a field of view from 3º to 17º in elevation and 336º in azimuth. Each telescope had a spherical mirror of 3.8 m$^2$ which was used to focus light on to a camera of 256 photomultipliers. Each photomultiplier had a hexagonal cross-section and subtended 1º x 1º on the sky. The second detector (HiRes-II) contained 42 telescopes, covering a greater elevation (3º to 31º) and 352º in azimuth. At closure in April 2006 4522, 4064 and 3460 hours of data had been accumulated with HiRes-I, HiRes-II and in the stereo mode respectively. Fuller details of the instrumentation are given in [76]. The analysis procedures follow those of Fly's Eye, outlined above, in many details.

In recent years a number of monocular spectra from HiRes-I and HiRes-II have been reported, most recently in [77] where a suppression of the spectrum above $4 \times 10^{19}$ eV was described. A stereo measurement has also been published [78] and is regarded by the group as their most accurate measurement although the energy range covered does not extend to such low energies and the observation time is shorter. The final spectra from HiRes-I and HiRes-II are shown in figure 10a while the stereo spectrum is in figure 10b.

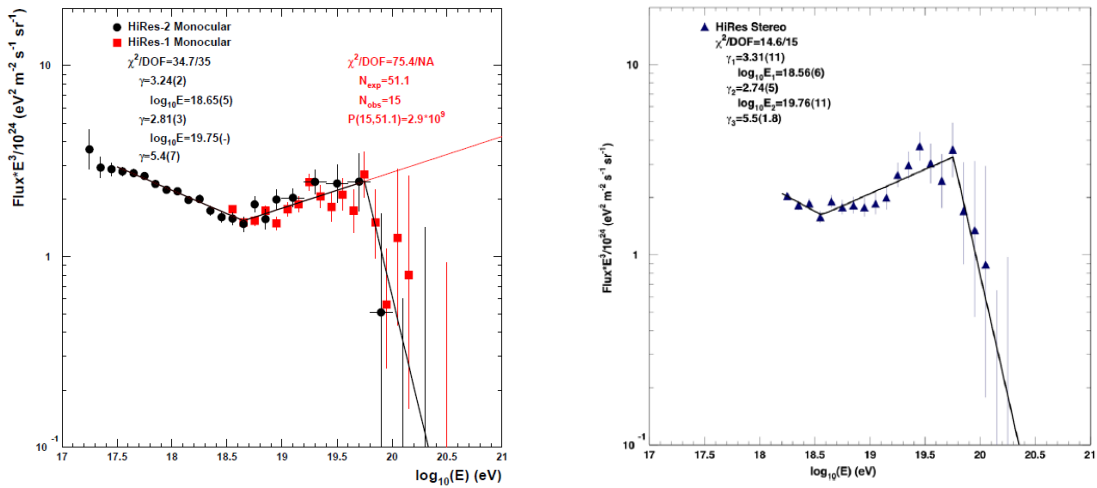

**Figure 10a and b:** Energy spectrum measured by the HiRes group using the two monocular instruments [77] and with the monocular instruments in stereo mode [78].

4.4 Energy Spectrum measurements from the Pierre Auger Observatory

The Pierre Auger Observatory is located near Malargüe, Argentina at a mean altitude of 1400 m above sea level (865 g cm$^{-2}$). 1600 water-Cherenkov detectors, each containing 12 tonnes of water are used to measure the response to the electrons, photons and muons of the showers. The surface array is laid out over 3000 km$^2$ on a triangular grid with a spacing of 1.5 km and is overlooked by 4 fluorescence detectors, each located on small hills at the edge of the array. Every fluorescence detector houses 6 telescopes, each with a field of view of 30º in azimuth and 1.5º to 30º in elevation. At each telescope light is focused onto a camera containing 440 hexagonal photomultipliers, each of 18 cm$^2$, at the focus of an 11 m$^2$ mirror. The design of the Observatory is described in [79]. Between 1 Jan 2004 (when data taking started) and 13 June 2008 (when construction was completed), the numbers of fluorescence telescopes increased from 6 to 24 and the number of surface detectors grew from 154 to 1600.

The time of the triggers at each water-Cherenkov detector is stamped using GPS and the direction of each event used for measurement of the energy spectrum is reconstructed with an angular accuracy of <1° [80]. The size parameter determined for each event is S(1000), the signal size in VEM estimated at 1000 m from the shower axis. Above $10^{19}$ eV the uncertainty in S(1000) is ~ 10%.

The longitudinal development of the showers in the atmosphere is measured using the fluorescence detectors with a Gaisser-Hillas function adopted to define the fit [81] with the energy derived with equation 6. To obtain the primary energy an estimate of the *missing energy* carried into the ground by muons and neutrinos is made based on assumptions about the mass of cosmic rays and of the appropriate hadronic model. The correction decreases as the energy increases (largely because of the reduced probability of decay of high-energy pions) and increases as the mass increases. Results of recent calculations [82] using the QGSJET II-3, QGSJET II-4, and EPOS 1.99 are shown in figure 11.

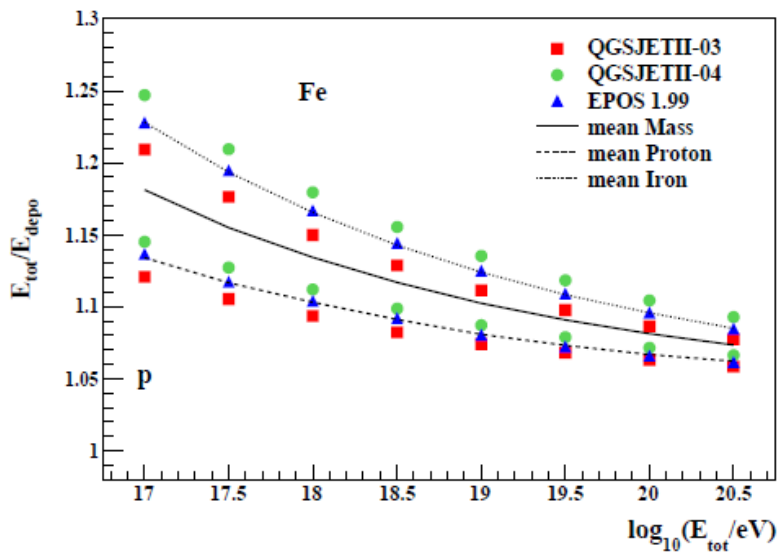

**Figure 11:** The ratio of the total energy ($E_{tot}$) to the energy deposited by the primary particle in the atmosphere ($E_{depo}$) is shown for several models and for proton and iron primaries [82].

Typical values of the correction, assuming a mean mass of 50% protons and 50% iron nuclei, are 14 and 9% at $10^{18}$ and $10^{20}$ eV respectively. At $10^{19}$ eV the average correction is ~ 9%, with a systematic uncertainty of ± 3%, corresponding to the extreme values from EPOS with Fe and from QGSJET-II with protons. Thus the derived energy estimates are systematically uncertain by these amounts. It is worth noting that the uncertainty in the energy estimate at $10^{19}$ eV is smaller than in several experiments near the knee region of the cosmic ray spectrum where fluorescence measurements are not possible and only model calculations are available. For the Auger work discussed here a composition of 50% proton and 50% iron has been adopted along with the QGSJETII model with the correction being dependent on energy.

An important feature of the design of the Auger Observatory was the introduction of the *hybrid technique* [83] as a new tool to study air-showers. The term is chosen to describe the method of recording fluorescence data coincident with the timing information from at least one surface detector.

An example of how the combination of times from surface detectors with the data from the fluorescence leads to an improvement in the angular accuracy is shown in figure 12 [84]. Two examples of the improvement in the essential parameters are shown. The gain obtained in the accuracy of the determination of $R_p$, the perpendicular distance from the fluorescence detector to the axis of the shower (equation 4) can be dramatic if the track in the fluorescence telescope is relatively short. The distance $R_p$ is important when determining the light emitted from the shower axis and when correcting for Rayleigh scattering and for absorption by aerosols. Reconstructions have been checked empirically using, *inter alia,* a centrally-positioned YAG laser of 7 mJ at 355 nm [85].

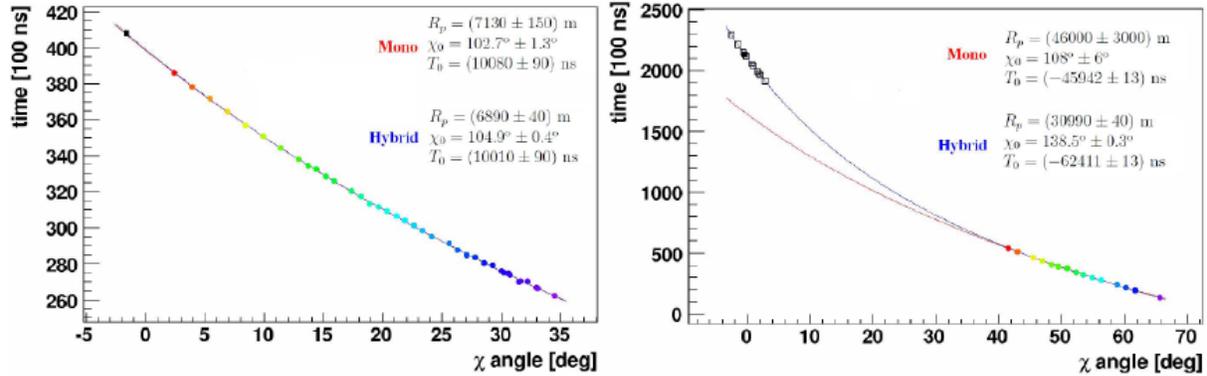

**Figure 12:** Two examples of events reconstructed using the hybrid technique. The addition of timing data from the surface detectors (points in the top left-hand ends of the lines in the two figures) alters the magnitude of the key parameters to be determined using equation 3. For certain geometries the times from the fluorescence detectors alone are not suitable for determining 3 parameters from this equation. The shower in the left-hand figure was at 35º from the zenith while that in the right-hand figure was from 73º [84].

The hybrid nature of the Auger Observatory enables the energy spectrum of primary cosmic rays to be found without a strong dependence on knowledge of the mass and hadronic interactions. This contrasts with what was necessary with systems that used only surface detectors such as Volcano Ranch, Haverah Park, SUGAR and AGASA where the use of models was essential for estimates of the primary energy.

The approach of the Auger Collaboration is to use a selected sample of hybrid events in which the energy can be estimated accurately with the fluorescence detectors. The resulting calibration curve then enables the energies of each of the much larger sample of events, for which there are only surface detector measurements, to be found. A recent calibration curve is shown in figure 13 [86]: the most energetic event in this sample has a total energy of 7.5 x $10^{19}$ eV. The parameter chosen to characterise the size of each SD event is the signal at 1000 m from the shower axes, S(1000), normalised to the median zenith angle of the events of 38°. The reasons for the choice of S(1000) as the *'ground parameter'* are described in [87]. The method used to combine events of different zenith angle is based on the constant integral intensity (CIC) method discussed above. The tailoring of this approach to the Auger data and the justification for normalisation to 38° are also described in [87]. Uncertainties in $S_{38°}$ and $E_{FD}$ are assigned to each event.

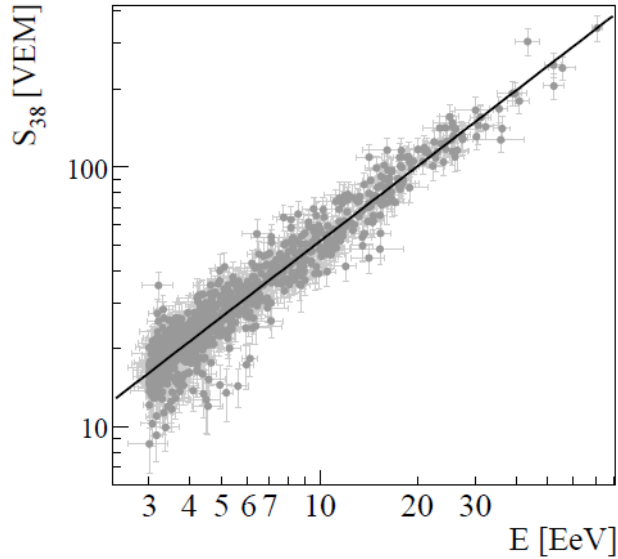

**Figure 13:** The calibration curve used to derive the signal measured by the surface detectors (y-axis) compared with the energy measured by the fluorescence detector for 839 selected hybrid events. The solid line represents the best fit to the data which extend to $7.5 \times 10^{19}$ eV [86].

For an event to be selected for inclusion in the energy spectrum, the detector in the shower that has the highest signal must be enclosed inside an *active* hexagon, defined as a hexagon in which all six of the surrounding surface detectors were operational at the time of the event. In this way it is guaranteed that the intersection of the axis of the shower with the ground (the shower core) is contained inside the array and therefore that the shower is sufficiently well-sampled to allow accurate measurement of S(1000) and the shower axis. From the analysis of hybrid events, and independently from Monte Carlo simulations, it is found that these selection criteria result in a 100% combined trigger and reconstruction efficiency for energies above $3 \times 10^{18}$ eV. The area over which the SD events fall and are recorded with 100% efficiency is found from the data and is independent of energy above $3 \times 10^{18}$ eV. The area does change over time but does not depend on the mass of the incoming particles. A major reason for change was the addition of water-Cherenkov detectors to the array while a further cause is the (rare) failures of individual detectors. The collecting area is calculated every second from the number of active hexagons.

To extend the measurement of the spectrum to lower energies, the Auger Collaboration has used hybrid events in which the shower has triggered at least one of the water-Cherenkov detectors in coincidence with a fluorescence telescope. The reconstruction of these events is superior to that with the surface detectors or with the fluorescence telescopes alone. For events above $10^{18}$ eV, the energy resolution is better than 6%. However, the calculation of the exposure is more complex and here it is necessary to use Monte Carlo calculations for which assumptions must be made about the primary mass composition. This leads to a systematic uncertainty in the aperture of about 8% at $10^{18}$ eV and around 1% at $10^{19}$ eV. Further details are given in [88].

A recent version of the Auger spectrum is shown in figure 14 where the hybrid spectrum and the spectrum deduced above $3 \times 10^{18}$ eV are shown both separately and combined [89]. There are ~ 64000 events above $3 \times 10^{18}$ eV of which ~5000 are above $10^{19}$ eV. These were accumulated in an exposure of 20905 km$^2$ sr yr: the hybrid spectrum is based on over 3000 events.

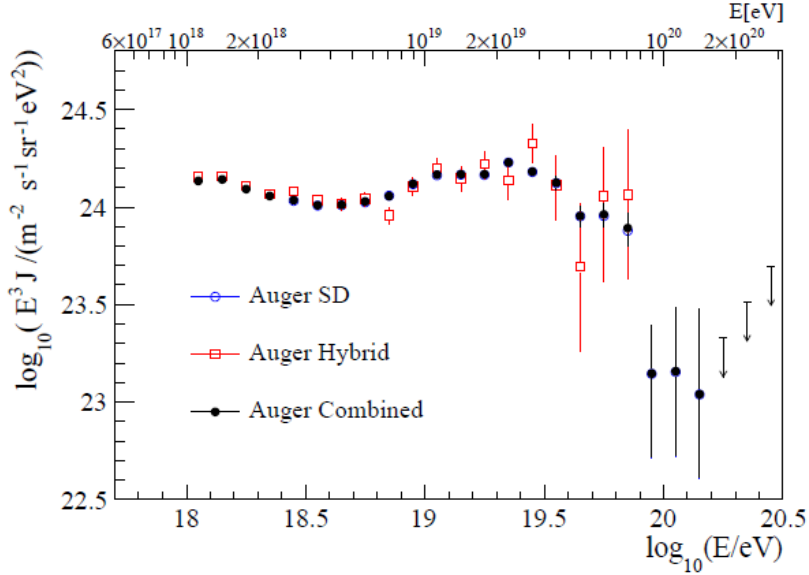

**Figure 14:** The energy spectrum measured with the Auger surface detectors using the calibration of figure 13. The spectrum from the surface detectors alone (above $3 \times 10^{18}$ eV) have been combined with the latest version of the hybrid spectrum [89].

The energy resolution of the data from the surface detectors improves from 20% at the lowest energies to about 10% at the highest energies and to make the best estimate of the true spectrum a forward-folding method has been applied.

4.5 Results on the energy spectrum from the Telescope Array
When AGASA and HiRes projects were nearing the end of operation, a collaboration consisting of key players from AGASA in Japan and the HiRes in US emerged to construct the Telescope Array (TA), the largest cosmic ray observatory ever built in the Northern hemisphere. Like the Auger Observatory, TA combines a large-area ground-array, similar to that at AGASA, with three fluorescence detectors overlooking it, one of which uses 14 telescopes from HiRes. TA is located in Millard County, Utah, at 39.3°N and 112.9°W at an altitude of 1400 m. The 507 surface detector stations are 3 m$^2$ scintillation counters spread approximately 700 km$^2$ with 1200 m spacing. The three fluorescence detectors contain 38 fluorescence telescopes.

The parameter used to characterise the size of events recorded with the surface detectors is S(800). Showers having zenith angles < 45° have been studied so far giving an aperture of 1100 km$^2$ sr above $6.3 \times 10^{18}$ eV where the detection efficiency is unity. Above this energy the shower direction is reconstructed with an accuracy of 1.5°. Fuller details can be found in [90]. A novel feature of TA is the inclusion of an electron linear accelerator that is intended to provide absolute calibration of the fluorescence detectors [91]. Data taking started in March 2008 and all 507 surface detectors were deployed by the following November. This impressively rapid operation involved the use of helicopters.

The construction of the TA energy spectrum [92] has been carried out in a manner similar to that introduced by the Auger Collaboration. The energy estimated with the fluorescence detectors is compared with the energy calculated from the value of S(800) measured under the assumption of proton primaries and the QGSJET-II-03 model of hadronic interactions. The energy found by the latter method is 27% greater than the FD energy and in figure 15 this has been taken into account in the scaling on the x-axis. This approach avoids the use of the CIC method but introduces uncertainties arising from the use of models and assumptions about the mass composition.

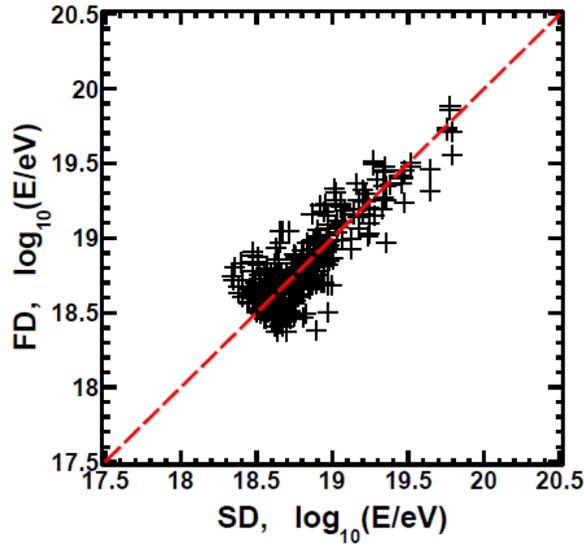

**Figure 15:** Energy comparison between the estimates of the energy made with the surface and fluorescence detectors of the Telescope Array after a 27% normalization has been applied to the surface array estimate made under the assumption of proton primaries and the QGSJET-II-03 model [92].

The energy spectrum derived from an exposure of 2640 km² sr yr is shown in figure 16. From ~ 6.3 x $10^{18}$ eV to ~ 1.6 x $10^{20}$ eV the efficiency is estimated using model calculations assuming proton primaries (see figure 17)..

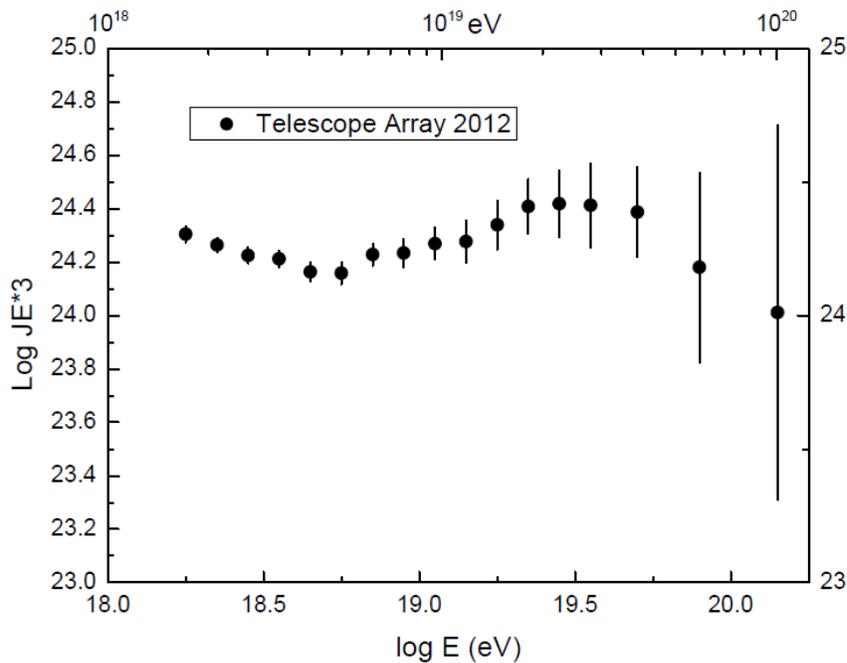

**Figure 16:** The energy spectrum of the Telescope Array [92].

5. Discussion of the different measurements of the energy spectrum

In this section a discussion of the differences between the energy spectra shown in figure 8 (AGASA [71]), figure 9 (Yakutsk [75]), figure 10a and b (HiRes [77 and 78]), figure 14 (Auger [89]) and figure 16 (TA [92]) is attempted.

A comparison of the slopes found in three different regions of the spectra, before the ankle (region I), between the ankle and the point of suppression (region II), and beyond the energy of suppression (region III) is made in table 1. Overall the agreement between the slopes is within the experimental uncertainties. The most striking differences are between the slope found for region II in the HiRes monocular measurement and the slopes measured by Auger and TA which are significantly smaller. The energy at which suppression sets in is lower in the Auger data than in those from the HiRes and TA.

A possible explanation of the first of these differences may lie in residual uncertainties in the apertures of the HiRes instruments that are quite strongly dependent on energy (figure 17). A relatively small change in this dependence would alter the spectrum slope. In the case of the Auger work, the aperture is independent of energy above $3 \times 10^{18}$ eV and is known to 3% [87] while for the Telescope Array the aperture is independent of energy above $\sim 6.3 \times 10^{18}$ eV (figure 17). At energies down to $\sim 1.6 \times 10^{18}$ eV, the efficiencies are estimated by HiRes and TA using model calculations under the assumption that the primary particles are protons.

|  | HiRes Mono [77] figure 10a | HiRes Stereo [78] figure 10b | Auger Combined [89] figure 14 | Telescope Array [92] figure 16 | Yakutsk Array [75] figure 9 | AGASA [71] Figure 8 |
|---|---|---|---|---|---|---|
| **Power Law before ankle, Region I** | **3.25 ± 0.01** | **3.31 ± 0.11** | **3.27 ± 0.02** | **3.33 ± 0.04** | **3.29 ± 0.17** | **3.21 ± 0.04** |
| **Power Law (intermediate), Region II** | **2.81 ± 0.03** | **2.74 ± 0.05** | **2.68 ± 0.01** | **2.68 ± 0.04** | **2.74 ± 0.20** | **2.69 ± 0.09** |
| **Power Law above suppression, Region III** | **5.1 ± 0.7** | **5.5 ± 1.8** | **4.2 ± 0.1** | **4.2 ± 0.7** | | |
| **log(E/eV) (ankle)** | **18.65± 0.05** | **18.56± 0.06** | **18.61± 0.01** | **18.69± 0.05** | **19.01± 0.01** | **18.95± 0.05** |
| **log (E/eV) (suppression)** | **19.75± 0.04** | **19.76± 0.11** | **19.41± 0.03** | **19.68± 0.1** | | |

**Table 1:** Fits to data from HiRes Monocular (figure 10a) and Stereo (figure 10b), the combined Auger data (figure 14), Telescope Array (figure 16), Yakutsk (figure 9) and AGASA (figure 8). Except for Yakutsk and AGASA, three power laws and two break points have been assumed in the fits. The values for AGASA have been computed by the author from the data used for figure 8 as no slopes were given in the associated paper [71]: there is no 'region III' in the cases of AGASA and Yakutsk.

The estimates of the energy at which the suppression occurs are slightly different and this is may be due to differences in the energy scales and/or a combination of other reasons including the choice of fit made to the data.

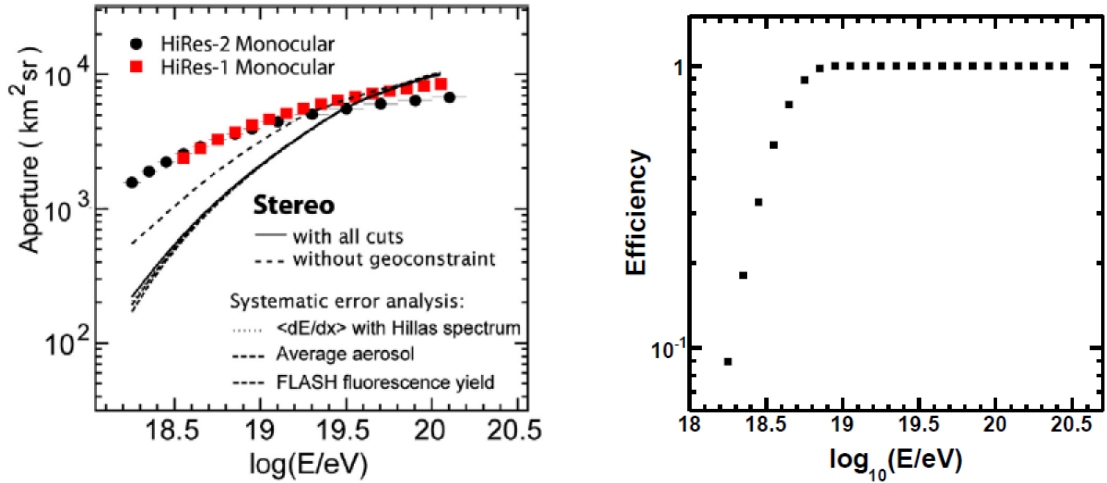

**Figure 17:** The aperture of the HiRes instruments (left plot) [78] and of the Telescope Array [92] as a function of energy.

In figures 8, 9, 10a and b, 14 and 16, the data are presented in the form of log $JE^3$ plotted against log E. This style of presentation is useful for emphasising features in spectra such as the ankle and the suppression and is a satisfactory presentation when the energy scale used for the x-axis is identical for the data sets shown, as in this set of figures. However in many review articles (including – regrettably – that by Nagano and myself [13]) it has been common to put data from different instruments on the same $JE^3$-style of plot.

I now consider this as bad practice for several reasons. Firstly, a logarithmic scale is very forgiving and can hide important differences between measurements; secondly, there may well be a significant uncertainty in the observed differential flux, $J_{obs}$, arising from the determination of the exposure and thirdly, the diagonal uncertainty associated with each point because of the uncertainty in the energy is rarely shown. I recommend, and adopt below, an alternative presentation in which the differential intensity at each energy is compared with the expected differential intensity from a *reference spectrum*: I find this to be a more informative style of presentation and what will be shown in several diagrams below is the *residual*, ($J_{obs} - J_{exp}$), normalised to $J_{exp}$, the expected intensity at a particular energy for the reference spectrum, calculated from a line of arbitrary slope passing through a measurement of high statistical weight. It is convenient to plot (($J_{obs}/J_{exp}$) – 1), the residual thus being expressed as a fraction of the expected intensity. This technique for comparing spectral data has been advocated previously [93] and has been commonly – but not invariably – adopted by the Auger Collaboration.

5.1 The reference spectrum
The reference spectrum has been chosen to have a slope 2.6 fitted to the flux measured by the Auger Collaboration [89] in the bin of ΔlogE = 0.1 at log (E/eV) = 18.65 which contains over 8500 events. The reference spectrum is thus:-

$$J = 3.56 \times 10^{16} \, E^{-2.6} \, m^{-2} \, sr^{-1} \, s^{-1} \, eV^{-1} \qquad (11)$$

The slope of the reference spectrum is quite arbitrary and has no physical significance.

5.2 Comparison of HiRes, Auger and TA measurements with the reference spectrum
In figure 19 the measurements from figures 10a, 10b, 14 and 16 are displayed as residuals calculated with respect to the reference spectrum. In figure 19a it is evident that there are differences between the HiResI and HiResII measurements that are large compared with the statistical uncertainties at some energies although the slopes and break points do agree within the statistical uncertainties. In figure 19b the HiRes Stereo data, the most accurate data set in terms of event reconstruction, show an intriguing feature at ~ $3 \times 10^{19}$ eV. This is also evident in figure 10b. Neither the differences between the HiResI and HiResII data, nor this feature, have been discussed by the HiRes group.

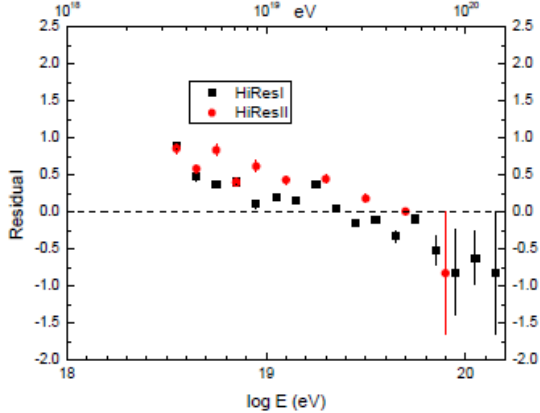
Figure 18a: Residuals for HiRes I and II

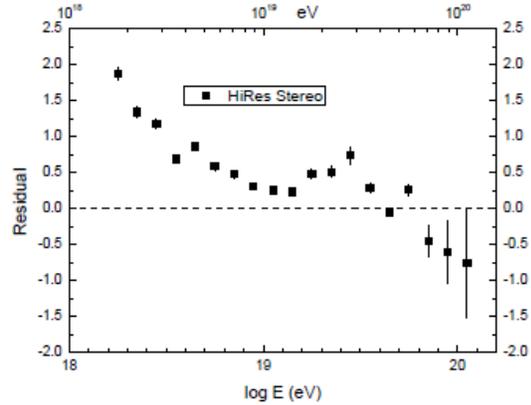
Figure 18b: Residuals for HiRes Stereo

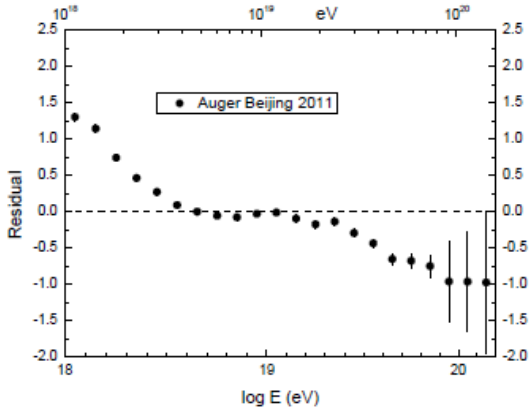
Figure 18c: Residuals for Auger Observatory

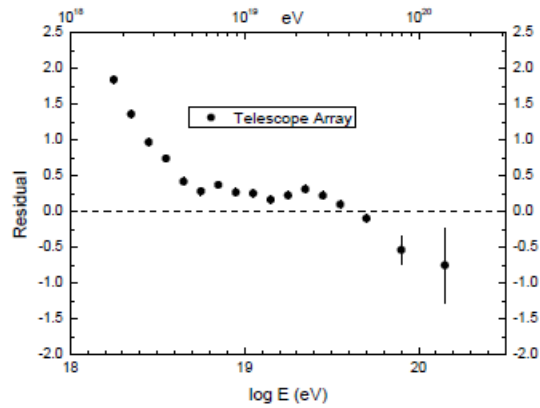
Figure 18d: Residuals for Telescope Array

Not surprisingly, in view of the choice of reference spectrum, the residuals from the Auger data fit well in region II (figure 18c).

The average of the residuals for the Telescope Array measurements is about + 25% in region II and there is also an intriguing suggestion of a bump at around the same energy as is seen in the HiRes stereo data. Note that this feature is scarcely evident in the $JE^3$ plot of figure 16 and no similar structure is seen in the Auger data. If the feature is real and the difference between Auger on the one hand and TA and HiRes on the other is established, it could be indicative of a difference between spectra from different hemispheres.

What does seem to be common to the spectra of figure 18 is evidence for a suppression of the flux of ultra-high-energy cosmic rays in the decade above $10^{19}$ eV, with the flattening of the spectrum (the ankle) at ~ $4 \times 10^{18}$ eV found in the HiRes, Auger and TA data. Indeed only the AGASA spectrum [71] fails to show evidence of flux suppression. The Yakutsk data, with the Cherenkov-light calibration, do not extend to a sufficiently high energy to make any statement. In the cases of AGASA and Yakutsk, the ankle is shown as being at ~ $10^{19}$ eV and ~ 6.3 x $10^{18}$ eV respectively.

In figure 19 the data on residuals displayed in figure 18 b, c and d have been brought together. As mentioned already, the small residuals for the Auger data arise because of the definition of the reference spectrum. There are some interesting differences between HiRes Stereo and TA as well as in the overall larger residuals as compared with Auger. The average residual for HiRes Stereo in the range 18.5 < log (E/eV) < 19.5 is (0.49 ± 0.06) as against (0.31 ± 0.03) for TA, i.e. the HiRes fluxes are 14% greater than those of TA although the important parameters used in conversion of the fluorescence signal to energy, such as the fluorescence yield and the correction for missing energy are identical. In the region above log (E/eV) = 18.85, and before the region of suppression, the average of the TA residuals is smaller at (0.24 ± 0.03). This may indicate that there is a systematic uncertainty at the lower energies still to be identified. The TA aperture is constant above log (E/eV) = 18.85 while the HiRes aperture is changing rather rapidly with energy (figure 17).

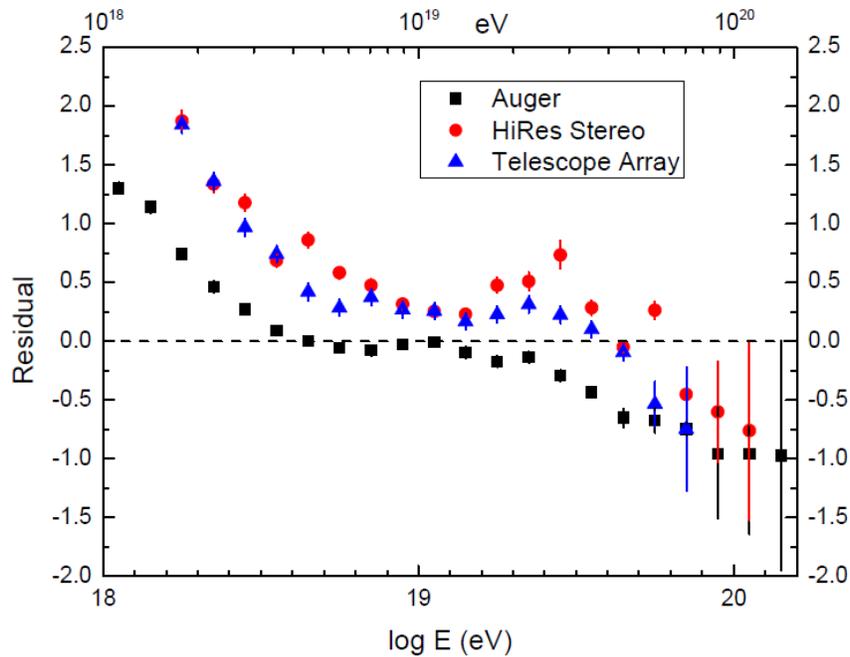

**Figure 19:** The residuals from the reference spectrum for the HiRes Stereo data, the Auger data and the data from the Telescope Array.

The HiRes and TA groups assume proton primaries in their calculations of the apertures, and thus in the determination of their spectra, so the validity of this assumption needs discussion. The HiRes group report that there is mass-sensitivity to the fluxes determined below $10^{18.4}$ eV. Also their calculations use the relatively old QGSjet01 model. It is not the purpose of this paper to review the evidence on primary mass at ultra high energies and the reader's attention is accordingly drawn to the review recently prepared by the Working Group formed from members of each of the Auger, HiRes, TA and Yakutsk collaborations on this topic [94] and endorsed by members of the Collaborations.

I will first give a brief summary of the report of the Working Group work. For some time, differences have been apparent between the inferences made about primary mass from measurements of the mean depth of maximum as a function of energy made by HiRes and by the Auger Collaboration. While the Auger group has argued that the mean mass becomes increasingly heavy above ~ 3 x $10^{18}$ eV [95] the HiRes group has claimed over many years that the mass composition is proton-dominated above a similar energy [96]. More recently the Telescope Array (TA) team has reported preliminary results [97] that appear to support the HiRes claims. After scrutinising the analysis procedures of the three groups, the Working Group concluded that there is no evidence to suggest that the apparent discrepancies are due to the different approaches but that there are important differences as to how the conclusions are drawn. In the case of the Auger Observatory, data from all zenith angles are combined directly and compared with model predictions: no differences are found between different zenith angle bands suggesting that the selection criteria are robust. In the case of HiRes and the Telescope Array the selection procedures are included as part of the Monte Carlo comparisons with the data. The Working Group has recognised that contrasting approaches make it impossible to compare the separate measurements of $X_{max}$ directly as has commonly been done. The Yakutsk method of deducing $X_{max}$ is to juxtapose the lateral distribution of Cherenkov light with model predictions and extract $X_{max}$ indirectly.

To compare the data sets, the Working Group have selected a particular model of the hadronic physics, Sibyll, and deduced the mean value of the logarithm of the atomic mass, <ln A>, measurement-by-measurement. Comparison between these alternatives has been made with a $\chi^2$ test to see whether the data from each observatory fit better the hypothesis of <ln A> being independent of energy or the hypothesis of a two-component fit with a break in energy at log (E/eV) = $10^{18.5}$. The Auger data are better fitted by the two-component model ($\chi^2$/ndf 134/6 (energy independence) and 7.4/9 (two-component fit)). For HiRes (4.4/7 and 1.2/6), TA (3.4/6 and 9.8/7) and Yakutsk (4.2/8 and 7.7/7) the three data sets are consistent with either model. What is clear is that around 3 times more events are needed from the Northern Hemisphere observatories to help resolve this issue. Furthermore data from other observations made with the surface arrays (muon densities in the case of Yakutsk and two independent approaches using the FADC traces from the Auger water-Cherenkov detectors), when treated in a similar manner, indicate that <lnA> is not constant with energy and that the mean mass becomes greater as the energy increases. Quite different systematic uncertainties are associated with the $X_{max}$ measurements and with those from the surface arrays. In my view the two-component model is favoured: unless there is a dramatic change in some features of hadronic interactions at energies beyond the LHC, the mean mass of the highest energy cosmic rays increases as the energy rises.

Further support for this position comes from an analysis that has been made using shower models enriched with the new information from the LHC. This analysis [98] shows that the depths of maxima are deeper in the atmosphere than had been computed previously. While the matter is not finally settled, my own view is that the balance of evidence favours the view that the mean mass becomes heavier as the energy increases. It is essential that the TA and HiRes data are reworked with models that incorporate LHC data.

Certainly at the higher energies, and possibly at lower energies if the particles are born in our galaxy, differences between spectra might occur because of near-linear propagation from nearby sources. It is known that the populations of possible sources are not identical in the parts of the universe viewed from the two hemispheres. In particular Centaurus A, a very active radio galaxy only ~ 3 Mpc from earth, might be expected to dominate in Auger data and it could be that the spectrum in the suppression region would be less steep than that observed from the Northern Hemisphere sites. However even with 1000 events above the suppression energy (currently ~ 600 have been recorded by the Auger Observatory) it would barely be possible to separate slopes of 4.2 and 4.7, particularly as there are currently no plans to have an array as large as the Pierre Auger Observatory in the Northern Hemisphere.

If the hypothesis of a galactic origin for all high-energy cosmic rays [6 ,7] was correct then it might be hard to exclude inequality of flux even at $10^{18}$ eV as there may be an important asymmetry in the galactic magnetic field. It would be a useful step in uncovering and understanding systematic differences if comparisons of the fluxes seen from overlapping parts of the sky were made. The fluxes measured by TA and Yakutsk, both scintillator arrays, could be compared in the Northern Hemisphere, while there is an overlap at more southerly declinations (-6º < δ < +16º) of the Auger and TA fields of view. These overlaps should be exploited with the aim of understanding some of the differences that presently exist and there is the intention between the collaborations to make this move.

There are several differences of detail as to how the estimates of the primary energies are derived at Auger, HiRes and TA. Better agreement between the results is obtained if the Auger energies are increased by 20%, an increase consistent with the systematic uncertainty of 22% reported by the Auger Collaboration [89]. However, there is no *a priori* reason to presume that the Auger energy scale should move up rather than the HiRes and TA scales be moved down. Furthermore it is not as straightforward to adjust the HiRes and TA spectra as it is to adjust that from Auger because of the energy dependence of the apertures (figure 17). Thus, in my view, it is premature to make more detailed comparisons.

There are other issues that need to be resolved before a final comparison between HiRes and Auger measurements can be made. One of these is the fluorescence yield to be adopted. The HiRes team has used the Bunner [65] measurement of the fluorescence spectrum combined with the accelerator results of Kakimoto et al [99] which give a direct measurement of the fluorescence yield. The Auger group have used a fluorescence spectrum measured by the AIRFLY collaboration [100] and combined this with the yields measured by Nagano et al. [101]. In a recent detailed review, Arqueros et al. [102] have shown that the yields adopted by the two groups at the dominant wavelength of 337 nm at p = 1013 hPa and T = 293 K are very similar, 5.4 photons MeV$^{-1}$ (HiRes) and 5.5 photons MeV$^{-1}$ respectively, despite the different input data. An absolute calibration from the AIRFLY collaboration is expected shortly but currently it does not appear that uncertainty over the fluorescence yield is a major issue.

A further important factor is the correction that must be made for missing energy. The HiRes and TA make a correction of 'about 10%' for the missing energy [78] although the correction is a function of energy of the primary and of the mass (see figure 11) and a constant correction does not seem appropriate. The Auger approach has been to assume a 50/50 mixture of proton and iron primaries and to use a correction that is energy dependent [82]. This is clearly an area that needs attention and is one of the factors that will need consideration if common-sky measurement is made.

5.3 Fits to the data and comparisons with predictions
Whilst the $\chi^2$-fit to two broken power laws is good for the HiRes measurements (35.1 for 35 degrees of freedom), the equivalent number for the Auger data is 37.8 with 16 degrees of freedom. As Nature is unlikely to deal in perfect power laws an alternative fit to the regions II and III was made for the Auger data using a function that is continuous through the region of suppression of the form:-

$$J(E; E > E_{ankle}) \propto \frac{E^{-\gamma_2}}{1 + \exp\left(\frac{\log_{10} E - \log_{10} E_{1/2}}{\log_{10} W_c}\right)} \quad (12)$$

where $\gamma_2$ is the slope of the spectrum in region II, $E_{1/2}$ is the energy at which the flux has fallen to one-half of the value of the power law extrapolation and $W_c$ defines the width of the transition region. The fitted $\chi^2 = 33.7$ for 16 degrees of freedom is only marginally improved with the relevant parameters being $\gamma_2 = (2.63 \pm 0.02)$, $\log(E_{1/2}/eV) = (19.63 \pm 0.02)$ and $\log(W_c/eV) = (0.15 \pm 0.02)$. Comparisons of the continuous fit to Auger data are shown in figure 20 where the results of calculations based on some phenomenology models are also displayed [103].

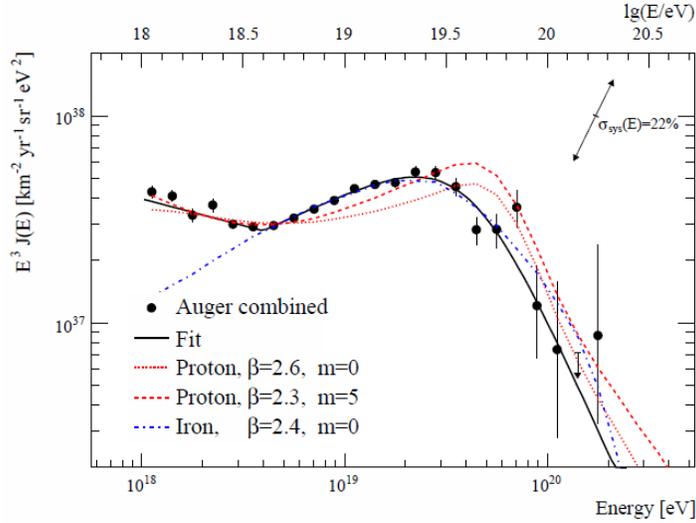

**Figure 20:** The fit to the Auger data to the function of equation 10 is displayed. The astrophysical models assume a pure composition of protons or iron, with a power-law injection spectrum ($E^{-\beta}$) and a maximum energy of $3 \times 10^{20}$ eV. The cosmological evolution of the source luminosity is $(z+1)^m$ [103].

In figure 20 a relatively limited range of input masses is considered while in figure 21 [18] the scenarios are extended to include the mass input as the galactic mixture and for other source evolutions.

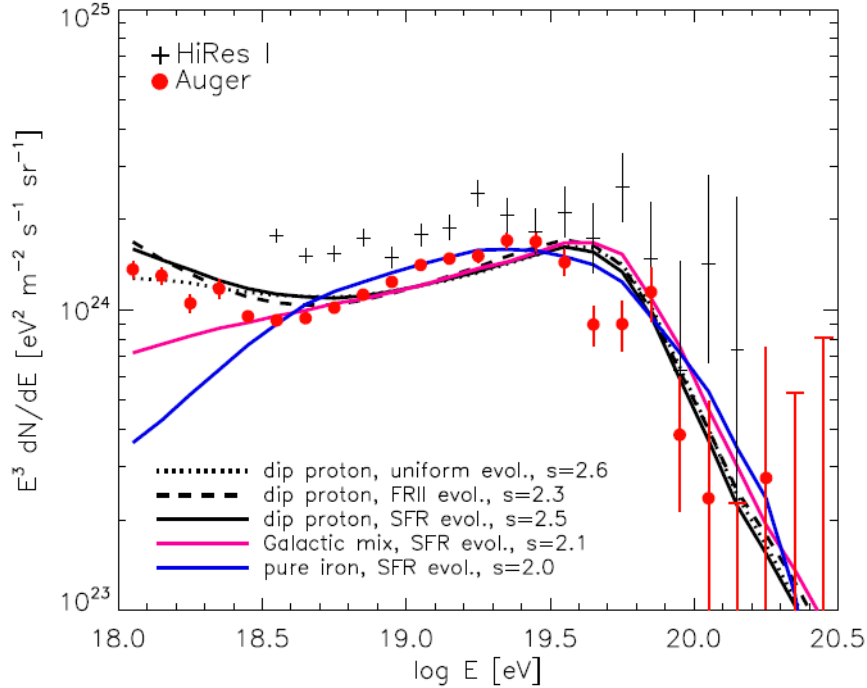

**Figure 21:** Spectral measurements by Auger and HiRes compared to simulated spectra for different models of the galactic to extragalactic transition and different injected chemical compositions and spectral indices, after [18].

In [104] similar studies have been further extended to include different maximum energies at the sources. It may be concluded that it is unlikely to be possible to distinguish unequivocally between the wide variety of different source inputs with the accuracy of data likely to be available in the next 10 years.

5.3 Summary of the position

Since the publication of papers reporting the observation of suppression of the spectra measured by the HiRes [77, 78] and Auger [87, 88] groups, it has been widely accepted that the suppression is a real effect and it is certainly highly significant in the statistical sense standing at 5 σ and over 20 σ respectively. While there are details of difference to be decided between these two measurements, and those from Telescope Array that show the same effect at 3.9 σ, the fact that they are relatively independent of assumptions about hadronic physics, by comparison with the AGASA measurement, make it likely that the suppression is a real effect.

6. Is the observed suppression evidence for the GZK effect?

An important question is whether the suppression in the energy spectrum at around $4 \times 10^{19}$ eV is evidence for the GZK-effect. While it tempting to believe that it has finally been observed, this may not in fact be the case.

Berezinsky has argued for many years that the highest energy cosmic rays are protons and that the suppression of the spectrum and the ankle can be explained as the imprints of propagation on the spectra of the particles at the sources. The suppression is then due to the γp process while the ankle is caused by pair-production on protons by the 2.7 K radiation, an effect first pointed out by Greisen [8] and Hillas [20]. Berezinsky et al. [105] have shown from their calculations that the *integral* intensity will fall to one-half of what it would have been had there been no GZK effect at an energy, $E_{1/2}$, such that log $E_{1/2}$(eV) = 19.76. The corresponding value derived from the HiRes monocular data [77] is log $E_{1/2}$(eV) = 19.73 ± 0.07. Matthiae has obtained log $E_{1/2}$(eV) = 19.8 using Auger data [106].

Some regard the agreement between the predictions and the measurements as evidence that the suppression is indeed the GZK effect and that the primary particles are all protons but I do not. Aside from the possibility of a numerical coincidence, there are substantial reasons for being cautious. Firstly the precision with which the energy is known absolutely, rather than relatively, is probably not yet better than about 20% (log ΔE = 0.08). Secondly, in my opinion, the arguments of the joint Working Group on UHECR mass composition, the later study in [98] and figure 20, certainly do not rule out a mixed composition at the highest energies, and the value of $E_{1/2}$ must surely depend on the mass composition of the particles injected at the source for which rather different spectra are expected to arise after propagation through the Universe (figures 2 and 21). Additionally a final spectrum, agreed by all collaborations, has still to be produced: all groups need to modify the allowances made for missing energy in the light of the LHC data and the uncertainty in the mass composition and should agree on the fluorescence yield to be adopted. The new hadronic models will lead to modifications in the aperture of the Telescope Array at energies below log (E/eV) = 18.8, and for HiRes at all energies, but in particular below $10^{18.4}$ eV where composition effects are important: such corrections can only be made by the groups involved. Finally an understanding of the structure apparent in the TA and HiRes stereo measurements at ~ $3 \times 10^{19}$ eV is required.

According to Berezinsky et al. [107] one interpretation of the Auger data may need what they call the 'the disappointing model' for the highest energy cosmic rays in which the break energy of $1.74 \times 10^{18}$ eV is seen as being indicative of the maximum energy ($4 - 6 \times 10^{19}$ eV) to which sources can accelerate protons with the highest energy reached in the source being Z times larger (~ 1 - $2 \times 10^{20}$ eV for iron nuclei). It is not clear to me why this model is described as 'disappointing' as I judge that it gives a rather elegant description of the measurements. It does, of course, lead to a reduction in the estimate of the flux of neutrinos and photons that might be expected at high energies - but this may be reality.

It is not going to be easy to resolve the ambiguity over interpretation of the suppression but help should come from studies of the arrival direction distribution of the highest energy events. In 2007 the Auger Collaboration reported a strong association of events with energies above 5.5 x $10^{19}$ eV with Active Galactic Nuclei [108]. This result was taken as being indicative of protons dominating at the highest energies. However a doubling of the data has not strengthened the effect and while a correlation with the local extragalactic matter distribution remains it may be that around 40 – 70% of the flux is isotropic [109]. However, further progress is severely hampered by the slow rate at which events accumulate above this energy, ~ 2 per month with the present Auger Observatory, and the possibility of determining the spectra from different sources, which would help to resolve matters, must wait until the JEM-EUSO mission [110] is launched and/or a World Observatory is built

7. Where have all the $10^{20}$ eV events gone?

In the review by Nagano and Watson [13] a tabulation of 14 events with energies reported as being > $10^{20}$ eV was given: Fly's Eye (1), Volcano Ranch (1), Haverah Park (4), Yakutsk (1) and AGASA (7). Since that publication, the estimates of the Haverah Park energies have been reduced by 30% as a consequence of re-estimating them using QGSJET98 to describe the shower development and a calculation based on GEANT to model better the response of the water-Cherenkov detectors [111]. However the AGASA and Yakutsk lists have increased to 11 and 4 respectively while the Fly's Eye and Volcano Ranch energies are unaltered. Thus there is a striking difference in the conclusions from the Auger, HiRes and TA measurements and that from, in particular, AGASA (compare figures 8, 10, 14 and 16).

For the few events claimed to have energies above $10^{20}$ eV by Auger (3), HiRes stereo (1) and TA (2) the integral flux is ~ (2.2 ± 0.9) x $10^{-4}$ km$^{-2}$ sr$^{-1}$yr$^{-1}$ while from 11 AGASA events the equivalent number is (6.35 ± 1.9) x $10^{-3}$ km$^{-2}$ sr$^{-1}$ yr$^{-1}$, a flux more than 25 times higher. I believe that it is inconceivable that the energy estimates by HiRes, Auger and TA can be out by a factor of 2 or that all devices have technical problems that have led to the most energetic events being lost.

However, in my view, there are no grounds for believing that the AGASA group made any errors in their measurement of the basic features of showers. The technique that they have used is extremely well-understood and was essentially the same as that introduced at Agassiz by the MIT team of Rossi in the 1950s and at Volcano Ranch in the 1960s and 1970s. There is an excellent understanding of the difficulties and systematic uncertainties that can arise with this technique and the scintillation counters used for the basic measurement are well-understood. The many groups that have operated ground-arrays have been in constant interaction and discussion with one another over the years. Nor is there any major uncertainty in corrections that have to be made for the fact that showers do not all come from the vertical direction although it is true that an accurate measurement is not possible, because of the small number of events at the highest energies. However one of the largest AGASA events [112] is rather vertical (22º) so the adjustment to 0º using the CIC method is small. However the energy of (2.1 ± 0.5 x $10^{20}$ eV) should probably be reduced by ~ 30% in line with the differences that are being found between the energies from fluorescence and model calibrations or more if the primary was a heavy nucleus. A detailed rebuttal of various criticisms of the AGASA measurements can be found in [71].

The famous Fly's Eye event of 3 x $10^{20}$ eV [68] remains the highest-energy cosmic ray ever observed.

8. Summary and Conclusions

The final words on the interpretation of the energy spectrum of the highest-energy cosmic rays remain to be written but it seems certain that the energy spectrum of cosmic rays steepens at about 3 - 4 x $10^{19}$ eV. This energy is close to that predicted by Greisen and by Zatsepin and Kuz'min as to where the flux spectrum would be sharply suppressed if the cosmic ray sources were distributed universally and had input spectra that continued to much higher energies. However in my view it remains unclear as to whether we are seeing the GZK effect or a feature associated with the physics of sources. That the spectral shape is not expected to be very different for proton or iron injections at source in some models means that measurement of the mass composition alone is unlikely to be sufficient to resolve this conundrum in the foreseeable future. The solution will surely come from better measurements of the arrival direction distributions at the highest energies and, to attain this, exposures even greater than will be achieved by the Auger Observatory in Argentina over the next 10 years must be made.

It was recognised soon after the first measurements of the energy spectrum by the Auger Collaboration in 2005 that the aperture was at least one order of magnitude too small. Given that Jim Cronin and I were often asked when pushing the project 'why do you want to make it so large?' this has been a huge disappointment and in part arose because earlier estimates of the UHECR flux had been too high. The challenge therefore is to find an economical method of increasing the aperture by around a factor of 10 or more. One approach is to take a fluorescence detector into space, as proposed by Linsley in the early 1980s, and plans are well-advanced through the JEM-EUSO instrument to do just that, attaching it to the Japanese module on the International Space Station in 2018 [110]. This extremely ambitious mission has three major goals. Firstly if neutrinos exist with energies above $10^{20}$ eV they will be detectable with JEM-EUSO: this would have a dramatic impact on astroparticle physics. Secondly it is anticipated that sources of UHECR will be identified. Thirdly the energy spectrum can be pushed to higher energies because of the large exposure. Where this instrument is more limited is in the assignment of atomic masses as the accuracy of the determination of $X_{max}$ in individual events cannot approach that of hybrid arrays on the ground.

Thus one can envisage a future as follows. On the ground the Pierre Auger Observatory and the Telescope Array will be enhanced, extending their measurement scope to embrace the study of features of hadronic interactions while continuing to enhance the astrophysical data. In space the JEM-EUSO mission will be used to identify neutrinos, extend the cosmic-ray spectrum and find sources of charged cosmic rays. The space observations will need to be followed up by a dedicated ground array of enormous size covering more than 30,000 $km^2$ and of a hybrid nature. The scale of this project can surely only be achieved through global cooperation perhaps to be realised in 8 to 10 years, but it is not too early to discuss plans, form collaborations and carry out R&D. I think of this *World Observatory* as being analogous to the LEP machine at CERN and the associated detectors that were built to study the W and Z properties in exquisite detail: JEM-EUSO might be thought of being analogous to the SPS. Of course costs must be kept as low as possible for a World Observatory to be supported and an innovative design for a cheaper fluorescence detector than has been possible in the past have been described [114]. The concept of a World Observatory will only become a reality if a team of dedicated people decide to devote a significant part of their careers to making it happen. Dynamic young scientists exist within the several collaborations: a few must step forward and lead the next stage. JEM-EUSO and a World Observatory must be seen as going hand-in-hand to the future: neither these two projects, nor the existing ones on the ground, are in competition. I believe that the Working Groups are an important first step in this adventure but the similar cooperation in hardware development is needed if progress is to be made. The rewarding results of the last two decades came as a consequence of the pioneers of the 1950s and 1960s [33]: I believe that we are now in excellent shape to make the similarly exciting advances.

Acknowledgments: I am most grateful to Misaki Fukushima (Telescope Array) and Anatole Ivanov (Yakutsk Array) for making available to me the data for their energy spectra. The continuing efforts, support and friendship of my Auger colleagues are gratefully acknowledged, as is the UK support to the Pierre Auger Observatory from the Science Technology and Facilities Council, and its predecessor, PPARC, over many years although, alas, it has now ended. I would also like to thank the Leverhulme Trust for financial support during 2011 and 2012. Etienne Parizot and Maria Monasor kindly provided figures 2 and 12 whilst I thank Carola Dobrigkeit for her heroic proof-reading. The views expressed in parts of this article are not necessarily in line with those of all, or even any, members of the Auger Collaboration. Two anonymous referees are thanked for their constructive criticisms and the Editorial Board for their angelic patience.